\documentclass[useAMS,usenatbib]{mn2e}

\usepackage{graphicx}
\usepackage{times}
\usepackage{natbib}
\usepackage{amsmath}

\renewcommand{\d}{\mathrm{d}}

\newcommand{\gtrsim}{\ga}
\newcommand{\lesssim}{\la}

\def\Omegab{{\Omega_{0,\rm b}}}

\def\Omegam{{\Omega_{0,\rm m}}}
\def\Omegal{{\Omega_{0,\rm \Lambda}}}

\title[Early objects with WDM]{
  The First Billion Years of a Warm Dark Matter Universe
}
\author[U.~Maio \& M.~Viel]{
Umberto~Maio$^{1,2}$\thanks{E-mail: maio@oats.inaf.it},
Matteo Viel$^{1,3}$\\
 ${}^1$ INAF -- Osservatorio Astronomico di Trieste, via G. Tiepolo, 11, 34131 Trieste (Italy)\\
 ${}^2$ Leibniz-Institut f\"ur Astrophysik (AIP), An der Sternwarte 16, 14482 Potsdam (Germany)\\
 ${}^3$ INFN/National Institute for Nuclear Physics, Via Valerio 2, I-34127 Trieste (Italy)\\
}

\begin{document}

\date{Accepted 2014 October 29. Received 2014 October 29; in original form 2014 September 24}
\pagerange{\pageref{firstpage}--\pageref{lastpage}}\pubyear{0}
\maketitle
\label{firstpage}

\begin{abstract}
We present results of cosmological N-body hydrodynamic chemistry simulations of primordial structure growth and evolution in a scenario with warm dark matter (WDM) having a mass of 3~keV (thermal relic) and compare with a model consisting of standard cold dark matter (CDM).
We focus on the high-redshift universe ($z>6$), where the structure formation process should better reflect the primordial (linear) differences in terms of matter power spectrum.
We find that early epochs can be exceptional probes of the dark-matter nature.
Non-linear WDM power spectra and mass functions are up to 2 dex lower than in CDM and show spreads of factor of a few persisting in the whole first Gyr.
Runaway molecular cooling in WDM haloes results severely inhibited because of the damping of power at large $k$ modes and hence cosmic (population III and II-I) star formation rate (SFR) is usually suppressed with respect to CDM predictions.
Luminous objects formed in a WDM background are very rare at $z>10$, due to the sparser and retarded evolution of early WDM mini-haloes during the dark ages and their lack can be fitted with a simple analytical formula depending only on magnitude and redshift.
Future high-$z$ observations of faint galaxies have the potential to discriminate between CDM and WDM scenarios by means of cosmic stellar mass density (SMD) and specific SFR, as well.
When compared to the effects of alternative cosmologies (e.g. non-Gaussian or dark-energy models) or of high-order corrections at large $z$ (e.g. primordial streaming motions or changes in the pristine IMF) the ones caused by WDM are definitely more dramatic.
\end{abstract}

\begin{keywords}
cosmology: theory - dark matter - early universe -- structure formation.
\end{keywords}


\section{Introduction}\label{Sect:introduction}


The first billion years is supposed to be the epoch when primordial structures could form by gravitational collapse and fragmentation of diffuse material \cite[][]{Hoyle1953,Peebles1969,Padmanabhan1993}.
The first bursts of star formation would then lead to the evolution and explosion of massive short-lived supernovae (SNe) that polluted the pristine Universe and contributed to heat up the intergalactic medium (IGM), playing a fundamental role for the progress of cosmic re-ionization. In this respect, numerical simulations have been useful to probe the infancy of the Universe and recent investigations have shed light on the characteristic masses, metallicities and star formation activity of primordial objects \cite[][]{deSouza2013}.
Despite typical properties of early proto-galaxies are significantly influenced by local effects and feedback mechanisms \cite[e.g.][]{BiffiMaio2013, deSouza2013arXiv, Wise2014}, some studies demonstrate possible dependencies on the background cosmological evolution, as well.
For example, non-Gaussianities, dark energy or primordial bulk flows could impact the first collapsing phases of pristine gas, at $z\gtrsim 10$
\cite[][]{Maio2006, Maio2011a, MaioIannuzzi2011,Pace2013arXiv}.
A closely related issue is about the important role that is played by the dark-matter component. An interesting possibility, advocated in order to solve small scale problems of the standard cold-dark-matter scenario, is that dark matter is made by warm particles that possess a non-negligible (albeit small) thermal velocity. This would affect the matter power spectrum producing a sharp decrease of power at relatively small scales \citep{AvilaReese:2000,Colin:2000,bode01}.
This aspect has profound implications for many astrophysical and cosmological observations, like weak lensing, clustering of galaxies, properties of haloes and their sub-haloes, strong lensing, intergalactic medium (IGM) structures, reionization, etc. \cite[see for a recent review e.g.][and references therein]{reviewWDM}.
The present tightest limits on cold dark matter coldness are placed by the IGM, namely the Lyman-$\alpha$ forest flux power spectrum as seen in absorption in quasar spectra, and give a 2$\sigma$ lower limit of $m_{\rm WDM} \gtrsim 3$~keV
\cite[][]{viel13}.
These values are larger than those that are typically chosen to reconcile local properties of the satellite populations of the Milky Way which are on the range $1.5-2\,\rm keV$ \cite[][]{maccio10,lovell12,lovell14,aba14}. However, we must caution that baryonic feedback can also play a crucial role in solving the small-scale cold-dark-matter crisis \cite[][]{brooks13} or having an impact on the matter power \cite[][]{Semboloni2011}.
Furthermore, if the IGM limits are applied, then there will be little room (if any) to disentangle between WDM and CDM in terms of low-redshift ($z<4$) structure formation \cite[][]{schneider14}.
Thus, it becomes apparent that there is an important new window that needs to be explored, also in view of future observational facilities: the high-redshift Universe.
\\
Previous studies of proto-star formation in pristine gas with WDM have shown that \cite[besides an expected suppression of clustering; ][]{yoshida03, gaotheuns07} the overall evolution and profiles of individual proto-stars remain qualitatively similar, although take place on different timescales \cite[][]{oshea06}.
These conclusions were based on pristine-cooling calculations that did not include star formation, stellar evolution, metal spreading from various generations of stars, feedback mechanisms, etc..
Further attempts to investigate the effects of WDM on cosmic structures have been presented relying on dark-matter merger trees coupled to recipes for baryon evolution \cite[e.g.][]{benson13,dayal14,bozek14}.
\\
Detailed numerical simulations including physical treatment for primordial star formation, stellar evolution, metal spreading and feedback effects to address the growth of visible objects have not been performed, yet (and they represent the goal of this work).
This is a sensitive limitation, because structure evolution in the first Gyrs could be a useful tool to place indirect constraints on the background cosmological model \cite[e.g.][]{Maio2011a, Maio2012}.
Additionally, the high-redshift regime has already been shown to be very promising in placing quantitative constraints on WDM models.
For example, high-$z$ Lyman-$\alpha$ forest data are more sensitive to the WDM cutoff than lower redshift data, since the flux power is probing more closely the primordial differences of the linear matter power spectrum, that are washed out by non-linear gravitational evolution \cite[][]{selsterile,viel05}.
Moreover, high redshift lensed $z\sim 10$ galaxies, can also put interesting (albeit weaker than IGM based) constraints on WDM scenarios in a way which is less dependent on astrophysical assumptions \cite[][]{pacucci14}.
Thus, pushing our studies towards primordial epochs might unveil possibly detectable WDM signatures.
\\
In this work, we present and discuss quantitative estimates of the baryon history as expected from numerical hydrodynamical chemistry simulations of structure formation in CDM and WDM models.
The paper is organized as follows:
the simulations performed are presented in Sect.~\ref{Sect:simulations}, 
while the results are discussed in Sect.~\ref{Sect:results}.
We finally conclude in Sect.~\ref{Sect:conclusions}.


\section{Simulations}\label{Sect:simulations}


The runs are performed by using the parallel N-body hydro code {\sc {P-Gadget-3}}, an updated version of the publicly available code Tree-Particle Mesh (PM) Smoothed Particle Hydrodynamics (SPH) {\sc {P-Gadget-2}} \cite[][]{Springel2005}, that, besides basic gravity and hydro, has been extend to include a detailed chemical treatment following $e^-$, H, H$^+$, H$^-$, He, He$^{+}$, He$^{++}$, H$_2$, H$_2^+$, D, D$^+$, HD, HeH$^+$ \cite[][]{Yoshida_et_al_2007, Maio2007, PM2012}, metal fine-structure transitions \cite[][]{Maio2007, Maio2009} and stellar evolution from different populations of stars \cite[][]{Tornatore2007, Maio2010, Maio2011b} according to corresponding metal yields (for He, C, N, O, Si, S, Fe, Mg, Ca, Ne, etc.) and lifetimes.
More precisely, pristine population III (popIII) stars are generated according to a top-heavy initial mass function (IMF) with slope $-2.35$ over the range $\rm [100, 500]\, M_\odot$.
If the metallicity, $Z$, of the star forming medium is higher than the critical value $Z_{\rm crit}=10^{-4}\,Z_\odot $ \cite[][]{Bromm2003,RS2003}, then the IMF is assumed to be Salpeter-like (population II-I, popII-I hereafter).
Both popIII and popII-I generations can enrich the early Universe with heavy elements, since the former can explode as pair-instability SNe (PISNe), possibly determining primordial supercollapsars \cite[][]{MaioBarkov2014}, and the latter as core-collapse SNe \cite[e.g.][and references therein]{HegerWoosley2002}.

\subsection{Implementation} \label{sect:implementation}

Our implementation follows standard gravity and hydro evolution together with a detailed non-equilibrium treatment, necessary for pristine regimes.
For each species ($e^-$, H, H$^+$, H$^-$, He, He$^{+}$, He$^{++}$, H$_2$, H$_2^+$, D, D$^+$, HD, HeH$^+$) with initial number density $n_i$ and temperature $T$, abundance variations within the time interval ${\d t}$ are given by:
\begin{equation}
\label{noneq_eq} 
\frac{\d n_i}{\d t}= \sum_p\sum_q k_{pq,i}(T) n_p n_q - \sum_l k_{li}(T) n_l  n_i,  
\end{equation}
where $ k_{pq,i}(T) $ is the (temperature-dependent) creation coefficient of species $i$ as a result of interactions of species $p$ and $q$, while $k_{li}(T)$ is the destruction coefficient due to interactions of species $i$ and $l$.
The complete list of reactions followed is given in Table~\ref{tab:reactions} \cite[see also][]{PM2012}.
\begin{table}
\begin{center}
\caption{ Reaction network}
\label{tab:reactions}
\begin{tabular}{lr}
\hline
\hline
Reactions & References for the rate coefficients\\
\hline
 	H    + e$^-$   $\rightarrow$ H$^{+}$  + 2e$^-$ & A97 / Y06 / M07\\
	H$^+$   + e$^-$  $\rightarrow$ H     + $\gamma$    & A97 / Y06 / M07\\
        H + $\gamma$ $\rightarrow$  H$^+$ + e$^-$  & A97 / Y06 / M07 \\
	He   + e$^-$   $\rightarrow$ He$^+$  + 2e$^-$    & A97 / Y06 / M07\\
	He$^+$  + e$^-$   $\rightarrow$ He   + $\gamma$     & A97 / Y06 / M07\\
        He + $\gamma$ $\rightarrow$  He$^{+}$ + e$^-$   & A97 / Y06 / M07 \\
	He$^+$  + e$^-$   $\rightarrow$ He$^{++}$ + 2e$^-$    & A97 / Y06 / M07\\
	He$^{++}$ + e$^-$   $\rightarrow$ He$^+$  + $\gamma$ & A97 / Y06 / M07\\
        He$^+$ + $\gamma$ $\rightarrow$  He$^{++}$ + e$^-$  & A97 / Y06 / M07 \\
	H    + e$^-$   $\rightarrow$ H$^-$    + $\gamma$     & GP98 / Y06 / M07\\
        H$^-$ + $\gamma$ $\rightarrow$  H + e$^-$  & A97 / Y06 / M07 \\
	H$^-$    + H  $\rightarrow$ H$_2$  + e$^-$       & GP98 / Y06 / M07\\
        H    + H$^+$ $\rightarrow$ H$_2$$^+$  + $\gamma$ & GP98 / Y06 / M07\\
        H$_2^+$ + $\gamma$ $\rightarrow$ 2 H$^+$ + e$^-$ & A97 / Y06 / M07\\
        H$_2^+$ + $\gamma$ $\rightarrow$  H + H$^+$  & A97 / Y06 / M07\\
        H$_2$$^+$  + H  $\rightarrow$ H$_2$  + H$^+$     & A97 / Y06 / M07\\
	H$_2$   + H   $\rightarrow$ 3H            & A97 / M07\\
	H$_2$   + H$^+$ $\rightarrow$ H$_2$$^+$  + H     & S04 / Y06 / M07\\
  	H$_2$   + e$^-$   $\rightarrow$ 2H   + e$^-$ & ST99 / GB03 / Y06 / M07\\
      	H$^-$    + e$^-$   $\rightarrow$ H    + 2e$^-$   &A97 / Y06 / M07\\
       	H$^-$    + H   $\rightarrow$ 2H    + e$^-$       & A97 / Y06 / M07\\
       	H$^-$    + H$^+$ $\rightarrow$ 2H                &P71 / GP98 / Y06 / M07\\
       	H$^-$    + H$^+$ $\rightarrow$ H$_2$$^+$  + e$^-$& SK87 / Y06 / M07\\
        H$_2$$^+$  + e$^-$   $\rightarrow$ 2H            &GP98 / Y06 / M07\\
        H$_2$$^+$  + H$^-$  $\rightarrow$ H    + H$_2$   &A97 / GP98 / Y06 / M07\\
        H$_2$ + $\gamma$ $\rightarrow$  H$_2^+$ +  e$^-$  & A97 / Y06 / M07 \\
        H$_2$ + $\gamma$ $\rightarrow$ 2 H & A97 / R01 / Y03 / M07 \\
        D    + H$_2$   $\rightarrow$   HD   + H     & WS02 / M07\\
        D$^+$  + H$_2$   $\rightarrow$   HD   + H$^+$  & WS02 / M07\\
        HD   + H   $\rightarrow$   D   + H$_2$         & SLP98 / M07\\
        HD   + H$^+$  $\rightarrow$   D$^+$  + H$_2$   & SLP98 / M07\\
        H$^+$  + D   $\rightarrow$   H    + D$^+$   & S02 / M07\\
        H    + D$^+$  $\rightarrow$   H$^+$  + D    & S02 / M07\\
        D$^+$  + e$^-$  $\rightarrow$   D + $\gamma$    & GP98 \\
        D  + $\gamma$  $\rightarrow$   D$^+$  + e$^-$  & GP98 \\
        He    + H$^+$  $\rightarrow$   HeH$^+$  + $\gamma$    & RD82/ GP98 / M07\\
        HeH$^+$    + H $\rightarrow$   He  + H$_2^+$    & KAH79 / GP98 / M07\\
        HeH$^+$    + $\gamma$ $\rightarrow$   He  + H$^+$    & RD82 / GP98 / M07\\
\hline
\hline
\end{tabular}
\end{center}
Notes:
$\gamma$ stands for photons;
P71~=~\cite{Peterson1971};
KAH79~=~\cite{KAH1979};
RD82~=~\cite{RD1982};
SK87~=~\cite{SK1987};
A97~=~\cite{Abel_et_al1997};
GP98~=~\cite{GP98};
SLP98~=~\cite{SLD_1998};
ST99~=~\cite{ST99};
R01~=~\cite{Ricotti2001};
WS02~=~\cite{Wang_Stancil_2002};
S02~=~\cite{Savin_2002};
GB03~=~\cite{GB03};
Y03~=~\cite{Yoshida2003};
S04~=~\cite{Savin_et_al2004};
Y06~=~\cite{Yoshida2006_astroph};
M07~=~\cite{Maio2007}.
\end{table}

\begin{figure}
\centering
\includegraphics[width=0.49\textwidth]{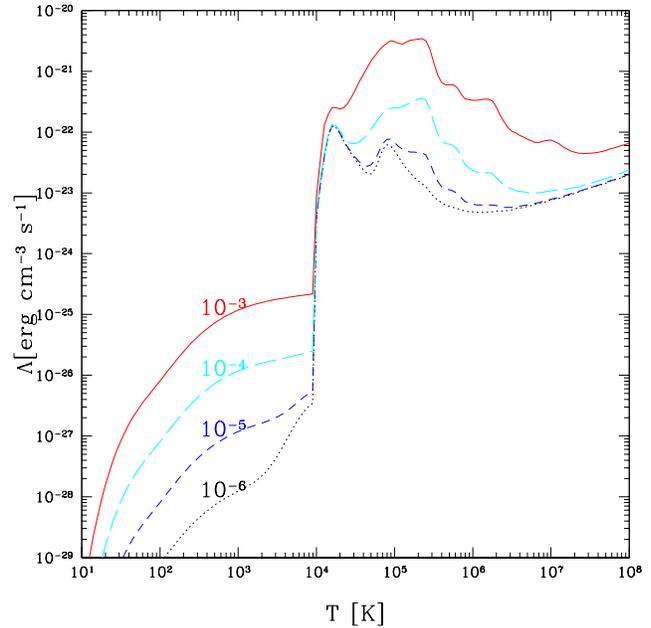}
\vspace{-0.5cm}
\caption{
  Total cooling due to hydrogen, helium, metals, H$_2$ and  HD molecules as function of temperature, for gas having a hydrogen number density of 1 $\rm cm^{-3}$.
The fraction of H$_2$ and HD are fixed to $10^{-5}$ and $10^{-8}$, respectively.
The labels in the plot refer to different amount of metals. See \citet{Maio2007} for further details.
}
\label{fig:cooling}
\end{figure}

Primordial gas parcels can cool mainly via H, He, H$_2$ and HD transitions.
In particular, at large temperatures, above $10^6\,\rm K$, radiative losses are driven by Bremsstrahlung (free electrons moving in the external field of protons and nuclides).
At temperatures around $\sim 10^4-10^5\,\rm $ the dominant coolants are H and He collisional excitations, while at temperatures $< 10^4\,\rm K$ H$_2$ is efficiently formed and molecular transitions can lower gas temperatures down to a few hundred Kelvin.
HD formation becomes then capable of bringing the temperature further down, below $\lesssim 10^2\,\rm K$ \cite[see e.g. Figure~2 in][]{Maio2007} thanks to its permanent electric dipole moment that determines smaller rotational energy gaps than~H$_2$.
\\
Early star formation episodes can suddenly enrich the medium of heavy elements produced in the inner cores of first stars.
Consequently, the numerous metal transitions will additionally contribute to gas cooling in the whole energy spectrum, from a few K to $\gtrsim 10^8\,\rm K$, according to the metal yields expected for different stars and after their corresponding lifetimes.
For the sake of clarity, we show the cooling function used in Figure~\ref{fig:cooling}.
This enters cooling time calculations and sets an upper limit on the simulation timestep.
In the implementation, we take into account the thermodynamical conditions of each individual gas element and assign the corresponding cooling energy according to its detailed chemical composition.

Star formation takes place in collapsing sites, where temperatures are usually $< 10^3\,\rm K$, at a rate (SFR) given by the local gas density and the local star formation timescale $\rm SFR \propto \rho/t_{sf}$ \cite[see more discussions in e.g.][]{Springel2005, Maio2009, Maio2013spin}.
Star formation episodes affects the surrounding medium via a number of feedback processes that regulate following collapsing events.
In particular, SN explosions produce energies as large as $\sim 10^{51}\,\rm erg$ and inject in the medium large amounts of entropy by shocking and heating (thermal feedback) the surrounding gas and by igniting galactic winds at velocities of a few hundreds km/s (kinetic feedback).
As a results of the feedback thermal effect, molecules can get dissociated and gas cooling momentaneously halted, while kinetic winds are mainly responsible for metal enrichment of the Universe.
These processes cause a {\it patchy} change in the cosmic chemical composition (chemical feedback) and induce a variation of the IMF features when metallicities reach values larger than $Z_{\rm crit}$.
The transition from the primordial (popIII) star formation regime to the metal enriched (popII-I) one is a consequence of the enhanced cooling capabilities of the gas in presence of heavy elements.
In such regions the medium can cool and fragment more efficiently thanks to the large variety of atomic metal transitions and hence smaller stellar masses than in pristine environments can be attained.
However, it must be noted that metallicity has a relevant role not only for the transition from the pristine popIII regime to the following popII-I, but also because it sustains gas cooling (by fine-structure lines) in the regions that have experienced molecule dissociation.
A deeper discussion about metal and molecular diagnostics in early (proto-)galaxies is found in the following Sect.~\ref{subsect:chemistry}).
\\
A popular issue is represented by the role of feedback from active galactic nuclei (AGN). We will not consider it in the present work, because AGN effects at the epochs explored here are definitely negligible.

The formation of primordial objects in the first Gyr is characterized by the build-up of a UV background at $z<10$ that can heat the IGM and the void regions.
Although included \cite[][]{HaardtMadau1996,HaardtMadau2001}, the UV background has little impacts on star formation, since this latter takes place in overdense shielded regions.
Large-scale processes, such as the UV background or even reionization can have implications on the thermal state of diffuse material, but they are not likely to alter dramatically the rapid runaway collapse of dense gas (which is clearly visible in many young star forming regions of the local {\it reionized} Universe).
Possible corrections due to unaccounted baryonic processes in both CDM and WDM will not interest the dark sector, but could cause systematic shifts in baryon properties.
However, the relative differences between the two scenarios would remain little affected.
As an example, the primordial stellar IMF is unknown and there are arguments both in favour of a high-mass range ($> 100 \,\rm M_\odot$) and in favour of a low/middle-mass range ($<40-100\,\rm M_\odot$), with consequent uncertainties on their metal yields and lifetimes.
Changes to the pristine IMF could cause temporal delays in the Population III history, as already investigated in literature \cite[e.g.][]{Maio2010, MaioBarkov2014}, but they would not affect the onset of star formation, mainly led by the background evolution \cite[see e.g.][Figure 4 and 5]{MaioIannuzzi2011}.
Similarly, primordial streaming velocities at initial conditions could surely induce a delayed collapse, but their impacts would be minor and degenerate with the background cosmological model \cite[Figure 6 in][]{MaioIannuzzi2011}.
We will see that, in general, WDM implications are neatly more important.

\subsection{Runs} \label{sect:runs}
Our simulations start at redshift $z=99$ for a $ L = 10\,\rm Mpc/{\it h}$ side box sampled with $2\times 512^3$ particles for dark matter and gas, respectively.
We perform two different simulations:  
$i)$ cold dark matter (CDM);  
$ii)$ warm dark matter (WDM) with a particle mass of 3~keV.
We decided to rely on such a WDM model for different reasons. 
First of all, this WDM particle is at $\sim 2\sigma$ confidence level consistent with all the known cosmological observables (including Lyman-$\alpha$ forest data); 
second of all, while being in agreement, this choice for the mass maximizes the differences with respect to the CDM model; 
thirdly, it is useful to quote results in terms of a thermal relic, because there is an exact correspondence between the mass and the free-streaming scale of the WDM cut-off -- this is no longer the case for other models like resonantly produced sterile neutrinos with leptonic asymmetry (for which the suppression is more similar to a step-like function and generally speaking less abrupt \cite[][]{Boyarsky:2008,boya09,Laine:2008}).
The cut-off in the linear matter power spectrum is chosen to be the one used by \cite{viel05} and is applied to the standard CDM power spectrum in order to compute the WDM one.
These two input spectra are then used to set up the initial conditions.
\\
In both scenarios matter, baryon and $\Lambda$ density parameters at the present time are chosen accordingly to recent cosmic microwave background data: $\Omegam=0.274$, $\Omegab=0.045$, $\Omegal=0.726$.
The expansion rate, $\rm H_0 \equiv 100\,{\it h}\, km/s/Mpc$, has $h=0.702$ and the power spectrum is normalized by imposing the mass variance within 8-$\rm Mpc/{\it h}$ radius sphere to be $\sigma_8=0.816$, with a slope $n_{\rm s}=0.968$.
The resulting dark matter and gas species have particle masses $\simeq 4.72 \times 10^5 \, \rm M_\odot/{\it h} $ and $ 9.47 \times 10^4 \, \rm M_\odot/{\it h} $.
Thanks to such resolution we are able to properly resolve typical Jeans masses in early mini-haloes ($\sim 10^6-10^8\,\rm M_\odot$) and the H$_2$ catastrophic cooling regime down to a density threshold of $\sim 10\,\rm cm^{-3}$.
We check smaller star formation thresholds of $1$ and $0.1 \,\rm cm^{-3}$ to study convergence and to highlight that below $\sim 1-10 \,\rm cm^{-3}$ there are severe overestimates for star formation rates at early times (see discussion in the next sections).
The properties of cosmic structures at different times are determined by employing a friend-of-friend and a Subfind algorithm \cite[see][]{Dolag2009}, by adopting a minimum number of 32 particles and a linking length of $0.2$ times the mean inter-particle separation.

\begin{figure}
  \includegraphics[width=0.48\textwidth]{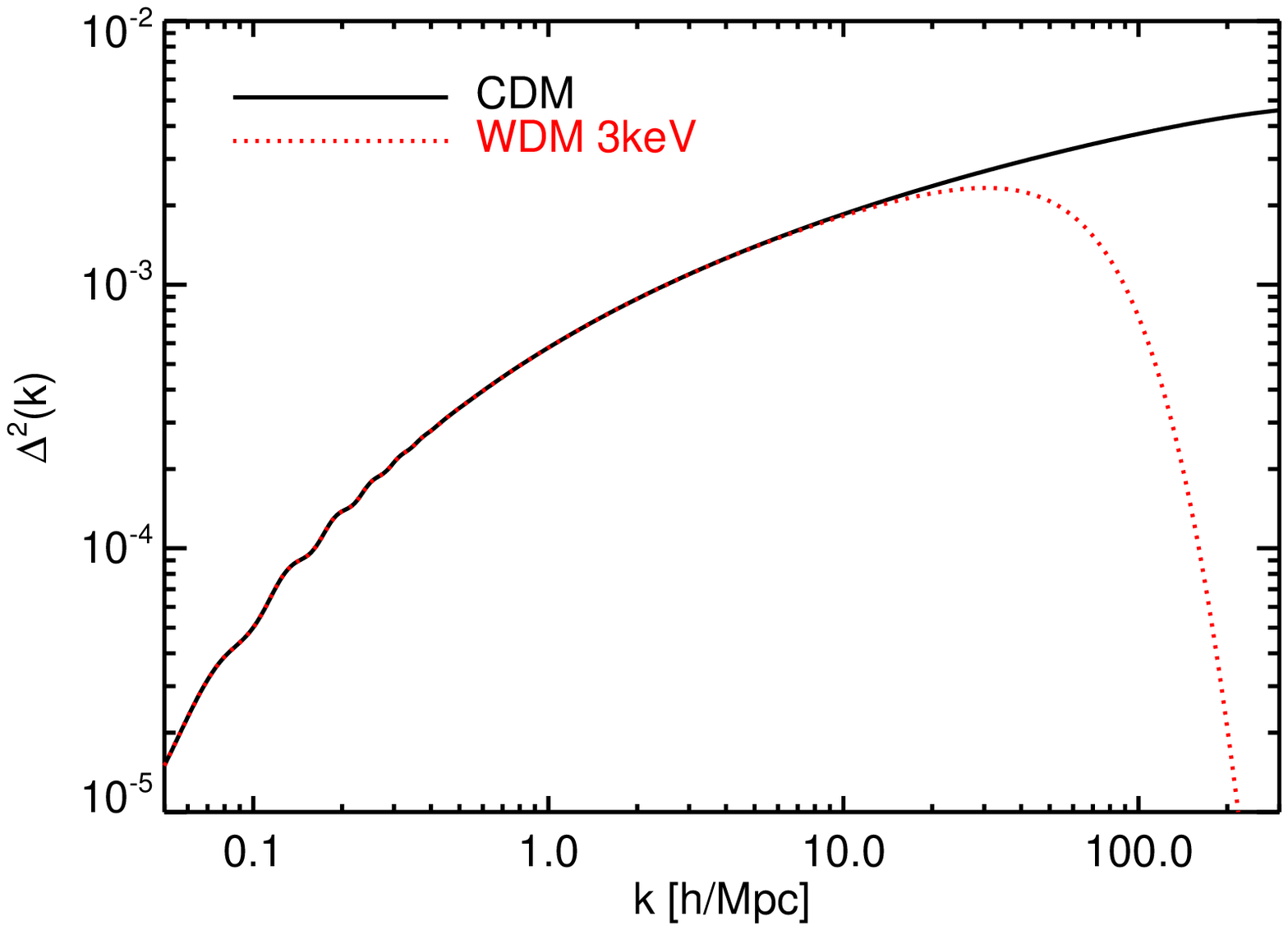}\\
  \vspace{-0.75cm}\\
  \includegraphics[width=0.48\textwidth]{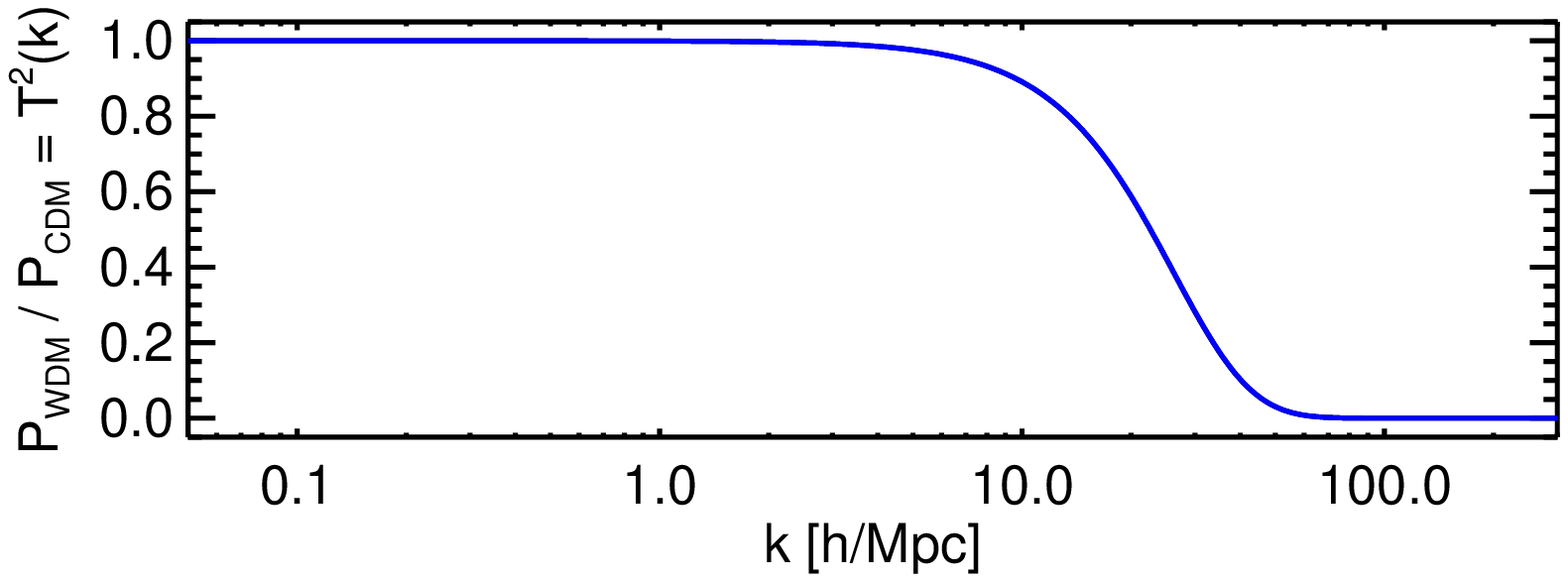}
  \vspace{-0.5cm}
  \caption[]{
    {\it Upper panel}.
    $\Delta^2$ computed from the linear matter power spectrum at $z=99$. The black continuous curve refers to the CDM case, while the red dotted one refers to the WDM case with a cut-off corresponding to a particle mass of a thermal relic of 3 keV.
    {\it Lower panel}. Corresponding WDM transfer function squared, representing the ratio between WDM and CDM power spectra.
  }
  \label{fig:Delta2}
\end{figure}

The Nyquist frequency for our grid of $N_{part} = 512$ particles a side corresponds to $k_{\rm Nyq} \equiv \pi \, N_{part} / L \simeq 160\,h$/Mpc.
The free-streaming scale (below which fluctuations are eliminated) associated to the chosen WDM particle mass is $ k_{\rm FS}\simeq 65\,h$/Mpc, while the scale at which the WDM suppression reaches the value of 50 per cent of the corresponding CDM power spectrum is $k_{1/2} \simeq 22.5 \,h$/Mpc, as visible in the lower panel of Figure~\ref{fig:Delta2} displaying the WDM transfer function \cite[see also][]{viel13}.
\\
In Figure \ref{fig:Delta2} (upper panel) we plot the input linear power spectrum at the initial time, $z=99$, for the two simulations in a-dimensional units ($\Delta^2$) in the upper panel, while in the lower
we show the WDM-induced suppression in the linear power spectra in terms of ratio of power spectra.
From this figure, it is clear that even if we are focussing on the high-redshift universe, there is a small amount of non-linearity present at the smallest scales that is due to the collapse of the first cosmic structures.
Another critical aspect that needs to be addressed is related to the presence of spurious numerical fragmentation in our WDM simulation.
This is a well known problem and it is discussed in \cite{wangwhite} and properly taken into account or quantified in several papers \cite[][]{lovell12, Schneider2012, Angulo13}.
We believe that our WDM simulation does not suffer of numerical fragmentation at the redshifts considered here.
Indeed, we do not observe any anomalous feature (i.e. steepening) in the mass function for small values of the mass of the haloes (we will discuss this referring to the left panels of Figure~\ref{fig:halomasses}).
First of all, this WDM particle is at $\sim 2\sigma$ confidence level consistent with all the known cosmological observables (including Lyman-$\alpha$ forest data).
Secondly, we sample our free-streaming scale, which is associated to a free-streaming mass of the early collapsing haloes of $\sim 2\times 10^8\, \rm M_{\odot}/{\it h}$, to a reasonably high accuracy.
Thirdly, we focus on relatively high redshift, where numerical fragmentation does not have a significant role, and we are using a large WDM mass that in turns means being closer to the CDM behaviour.
Moreover, we note that recent high-resolution dark-matter simulations \cite[][]{Schultz2014} do not show any numerical fragmentation for a mass resolution which is a factor 15 lower than ours and for a value of the thermal mass ($m_{\rm WDM}=2.6\,\rm keV$) that is very similar to ours.
All the arguments above suggest that our structure formation process is not going to be affected by spurious fragmentation.
For consistency, we will explicitly check these issues as well as their dependences on resolution in the following sections.

\section{Results}\label{Sect:results}
\begin{figure*}
\centering
{\bf\large  CDM   \hspace{0.7\textwidth} WDM}\\
\includegraphics[width=0.95\textwidth]{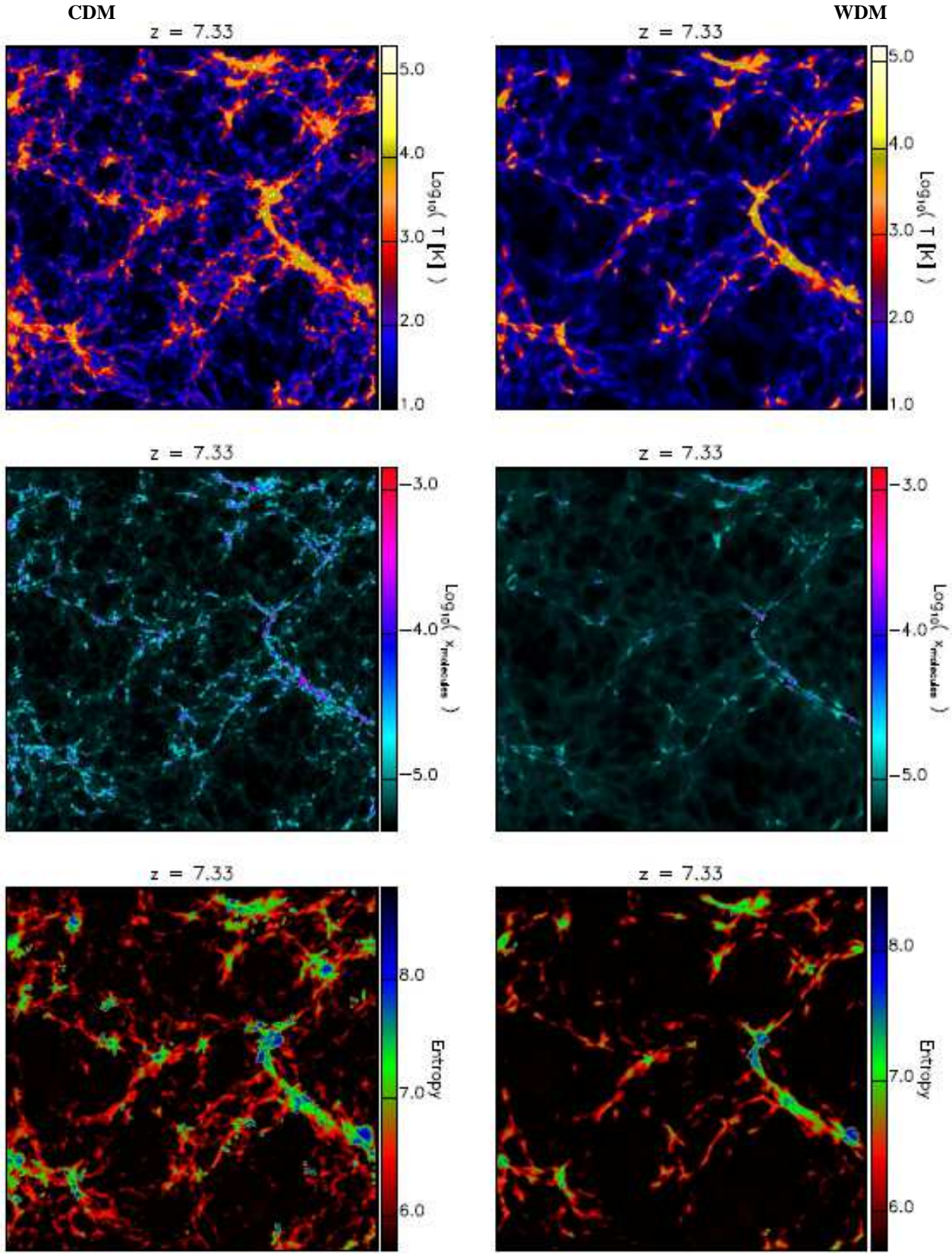}\\
\caption[]{
  Maps of cosmic structure properties for CDM (left) and WDM (right) scenarios. Projections are performed for a 0.1~Mpc/{\it h} (comoving) thick slice passing through the middle of the box for mass-weighted temperature (top), molecular fractions (center) and entropy (bottom). The linear size of the box is 10 comoving Mpc$/h$.
}
\label{fig:maps}
\end{figure*}
A visual representation of the numerical results is given in the maps of Figure~\ref{fig:maps}, where temperature, $T$ (top), molecular abundance, $x_{\rm molecules}$ (center), and entropy (bottom) are shown both for the CDM (left panels) and for WDM (right panels) model.
From a first inspection at the temperature maps it appears that in the CDM model there are more structures and substructures than in the WDM model, where cosmic objects and filaments are smoother and less clustered.
This is a direct consequence of the different input power spectra at large $k$ values, that determine a suppression of WDM structures with respect to the CDM case.
Correspondingly, there are numerous dense molecular-rich sites (see e.g. the high-$x_{\rm molecules}$ pixels in pink and red in the molecular-fraction maps) undergoing runaway cooling in the CDM scenario and a paucity of such regions in the WDM case.
The ongoing star formation and feedback mechanisms influence both gas density and pressure and, in fact, the more intense activity and the more advanced stages of collapsing material in CDM with respect to WDM is well visible from the entropy maps.
Particularly clear are the effects of shocks, winds and thermal heating (large entropy -- blue in the maps) originating in the cold dense cores (low entropy -- red and green in the maps) as a results of gas cooling and star formation.
These considerations apply at any redshift and characterize growth and evolution of the infant cosmological structures.
\\
In the following, more quantitative results are presented and their implications discussed.

\subsection{Cosmological evolution} \label{subsec:evolution}
\begin{figure}
\includegraphics[width=0.475\textwidth]{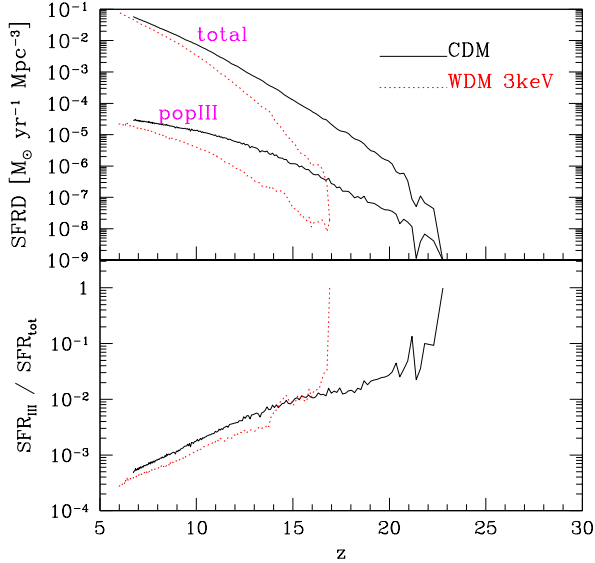}
\vspace{-1cm}
\caption[]{
 Total star formation rate densities and population III star forming gas in the CDM and the WDM (3 keV) models, as indicated by the labels for both the CDM and in the WDM (3 keV) scenarios. Corresponding popIII contributions to the cosmic star formation rate are shown in the bottom panel.
}
\label{fig:sfr}
\end{figure}

To have a first quantitative estimate of the implications of WDM models with respect to CDM ones, we plot in the upper panel of Figure~\ref{fig:sfr} the corresponding cosmic star formation rate densities~(SFRDs).
\begin{figure}
\includegraphics[width=0.45\textwidth]{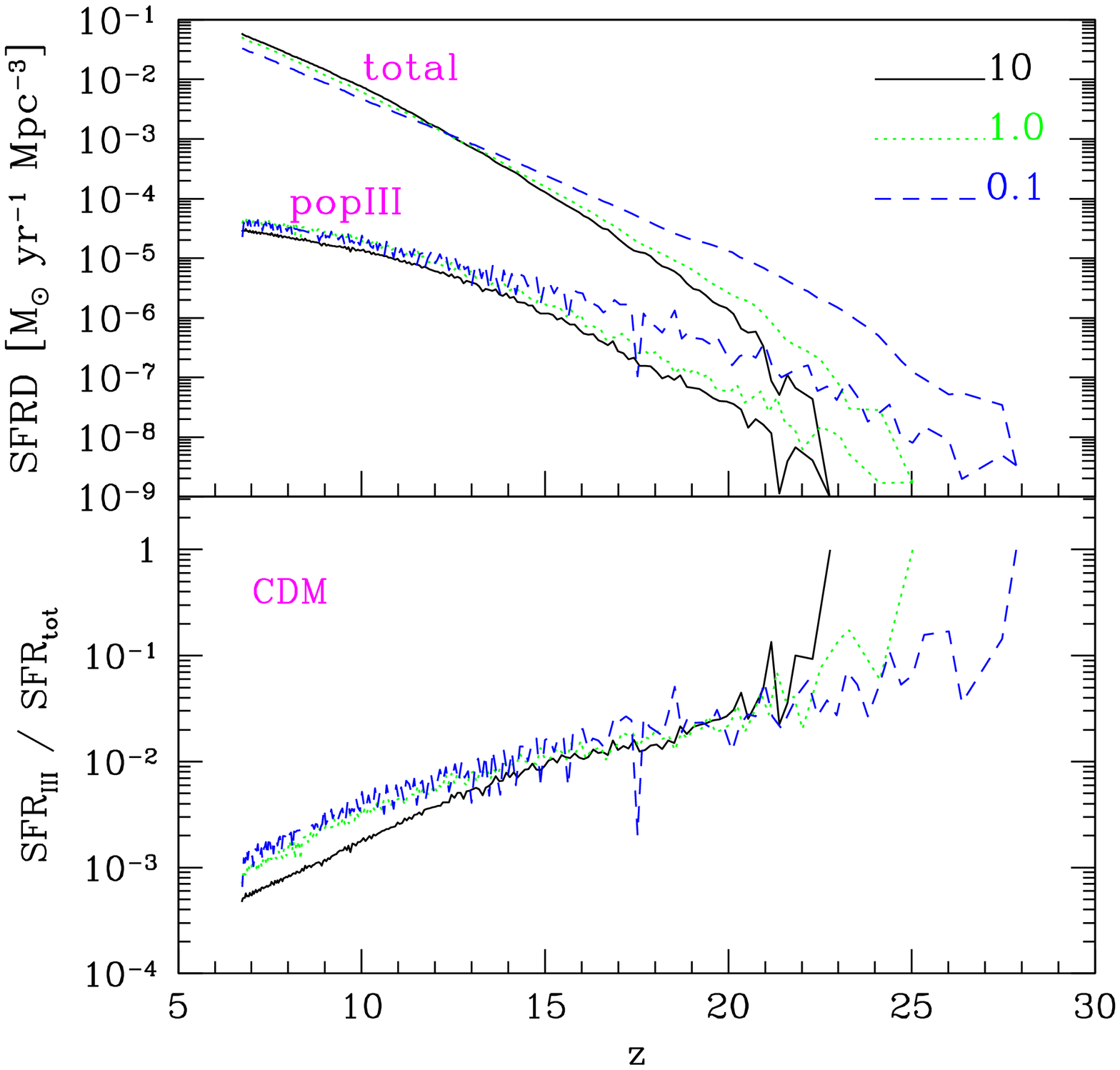}
\vspace{-0.5cm}\\
\includegraphics[width=0.45\textwidth]{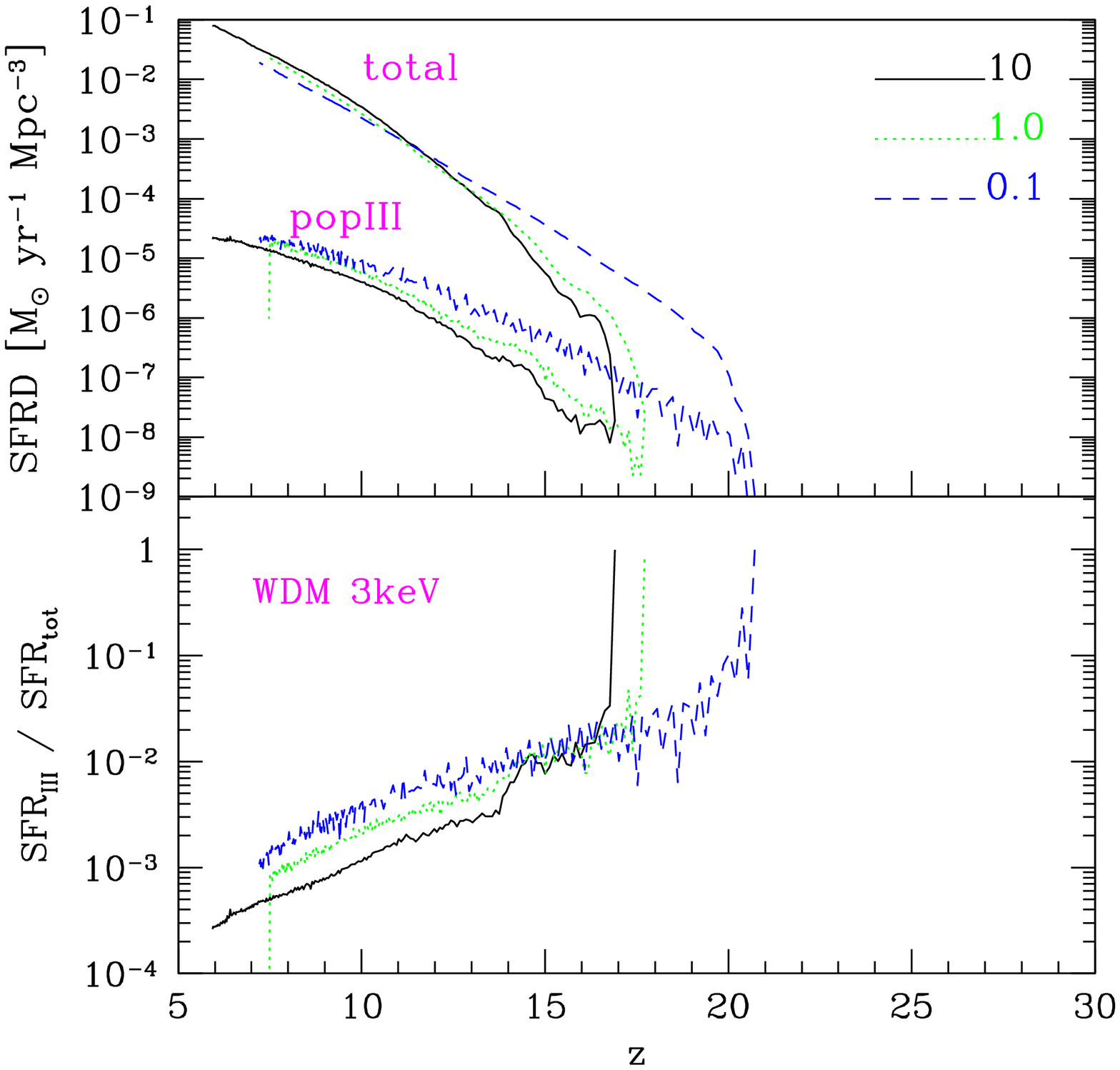}
\vspace{-0.5cm}
\caption[]{
  Total star formation rate densities and popIII contributions for runs with density thresholds of 10 (solid lines), 1 (dotted lines) and 0.1 $\,\rm cm^{-3}$ (dashed) both in the CDM and in the WDM (3 keV) scenarios.
}
\label{fig:sfr_convergence}
\end{figure}
\begin{figure}
\centering
\includegraphics[width=0.45\textwidth]{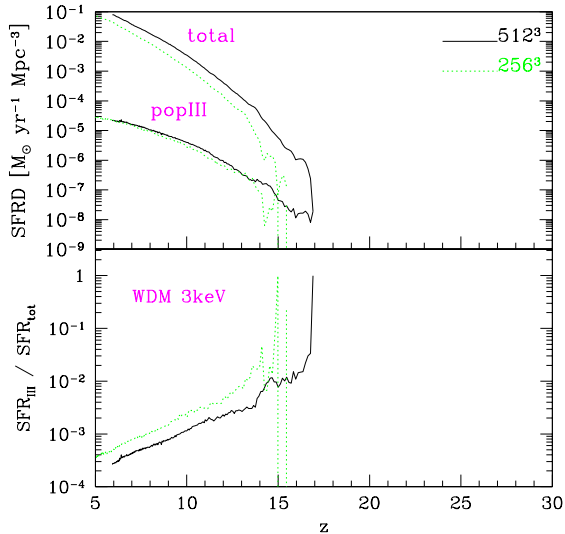}
\vspace{-0.5cm}
\caption[]{
  Total star formation rate densities (upper panel) and corresponding popIII contributions (lower panel) for the WDM model run at resolutions of $512^3$ and $256^3$ particles per gas and dark species in a $\rm 10\, Mpc/{\it h}$ side box.
}
\label{fig:sfrresolution}
\end{figure}

In both cases the popIII SFRD (upper panel) and its contribution to the total one (lower panel) are reported, as well.
The most striking effect of WDM results to be a dramatic drop of star formation activity in the whole first billion years.
Indeed, while primordial gas collapse and the following onset of star formation take place by $z\simeq 23$ in the CDM case, they are delayed by $\Delta z\sim 6$ (i.e. about 0.1 Gyr)\footnote{
The cosmic time, $t(z)$, at these redshifts is $t(17)=0.233\,\rm Gyr$ and $t(23) = 0.151\,\rm Gyr$.
}
in the WDM model.
Both trends are quite steep, but they differ significantly at $z\sim 10-15$.
At $z \gtrsim 10$ the WDM SFRD is always a few orders of magnitudes lower than the CDM SFRD and convergent values are retrieved only around $z\sim 6-10$, with $\rm SFRD\rm \sim 10^{-1}\, M_\odot yr^{-1} Mpc^{-3}$.
The evolution of popIII SFR, $\rm SFR_{III}$, with WDM features a similar delay and the contribution from the popIII regime steeply decreases from unity down to $\sim 10^{-2}$ at $z\sim 15$ for WDM, while the trend is much smoother for CDM.
These behaviours can be understood as the result of the lack of power at small scales in the WDM model compared to the CDM model.
This inhibits primordial collapse of small WDM structures and determines a prolonged phase of the dark ages during which cosmic gas cannot reach high densities and hence cannot cool efficiently.
Only at later times, when structures have grown significantly, there is a partial catch-up of the WDM star formation activity (both popIII and popII-I) that rapidly converts into stars the gas that has meanwhile accumulated in larger haloes (see also the specific SFRs later).
In the CDM model, instead, the SFRD is more gradual, because there is no cut-off in the CDM power spectrum and thus all the haloes (even the small primordial ones) have potentially the power to grow, collapse and host some star formation.
After the first episodes of star formation the popIII contribution follows roughly similar, although subdominant, evolutions for both WDM and CDM.
This is due to the fact that metal pollution is the main driver of stellar population transition \cite[e.g.][]{Tornatore2007, Maio2010, Wise2014} and the high yields of pristine stars determine a rapid enrichment of the surrounding regions turning most of them in possible popII-I star forming sites. 
The residual popIII activity is due to unpolluted material in the outskirts of enriched haloes or to newly born pristine haloes forming at a rate dictated by the background cosmological model.
We stress that in the scenario with CDM, metal pollution starts earlier, when the hosting haloes are usually smaller and can witness popIII star formation at $z > 15$.
On the contrary, WDM $\rm SFR_{III}$ contribution immediately drops down, because enrichment takes place suddenly, as soon as the (fewer) larger haloes reach the collapse scales.
Thus, these first events are hosted in WDM haloes at $z\sim 15$ that are typically larger (by a factor of $\sim 7$)  than the CDM counterparts at $z\gtrsim 20$.
This means that WDM star forming sites can locally convert slightly more gas into stars and can experience more chemical feedback, which limits the possibilities for subsequent popIII episodes.
This translates into lower $\rm SFR_{III}$ amounts by almost a factor 2 between WDM and CDM, at $z \sim 5-15$.
\\
Interestingly, Figure~\ref{fig:sfr} seems to suggest that the epoch of reionization ($z\gtrsim 6$) is strongly interested by these processes and the resulting 21-cm signal would be quite sensitive to the underlying dark-matter scenario.
Therefore, WDM models might be further constrained by reionization data \cite[e.g.][]{Fan2006, Schultz2014} and current or future missions aiming at studies of the cosmic dawn, such as
LOFAR \cite[][]{vanHaarlem2013LOFAR},
PAPER \cite[][]{Parsons2010PAPER},
MWA \cite[][]{Tingay2013SKAprecursoMWA},
SKA \cite[][]{CarilliRawlings2004, Dewdney2013SKA},
\cite[it is not obvious whether JWST could help; e.g.][]{dayal14}.

Our findings for CDM and WDM models are confirmed even when considering different density thresholds for star formation.
Figure~\ref{fig:sfr_convergence} demonstrates that the qualitative behaviours are preserved, none the less there are some systematic differences for the cases referring to thresholds of 0.1, 1, 10~$\,\rm cm^{-3}$, both for CDM and WDM.
We remind here that when primordial gas falls into the growing dark-matter potential wells it gets shock-heated at $\sim 10^4\,\rm K$ and stays around or below such value because of efficient atomic H cooling (see also Figure~\ref{fig:cooling}). In such regime molecule formation can take place even though this process is initially slow, due to relatively low densities. This determines a thermal phase where temperatures stay around $\sim 10^3-10^4\,\rm K$ and densities $\sim 1\,\rm cm^{-3}$ ({\it loitering regime}). Above such density value, the increase of collisional events boosts H$_2$ formation, the consequent runaway cooling down to $\lesssim 10^2\,\rm K$ and the following star formation\footnote{See Figure~2 in \cite{Maio2011b} for a schematic representation.}.
Models with smaller density thresholds predict, {\it cet. par.}, earlier onsets, because they would form stars by converting material that is still underdense, hot or in the loitering regime.
This issue is particularly relevant for threshold values $\lesssim 1\,\rm cm^{-3}$.
In such cases primordial gas cannot form large amounts of molecules and the resulting baryonic evolution would likely suffer of numerical problems.
Given these considerations, in the next we will focus mainly on the cases with 10~$\,\rm cm^{-3}$ threshold and refer the interested reader to our previous studies and numerical tests in the literature \cite[e.g.][and references therein]{Maio2010, Maio2011b}.

We close this section by addressing the effects of numerical resolution on the cosmic SFR.
We present the results in Figure~\ref{fig:sfrresolution}, where we consider the WDM scenario in $\rm 10\,Mpc/{\it h}$ side boxes sampled at $z=99$ both with $512^3$ and $256^3$ particles for gas and dark matter. 
\\
Total and popIII SFRDs feature similar evolution, starting at $z\sim 15$ and regularly increasing for the whole first billion years.
Never the less, some discrepancies due to resolution are visible.
The onset is very noisy in the low-resolution run and the following trend is systematically lower then the high-resolution one.
These are consequences of the more coarse sampling of primordial structures and of the lack of small (unresolved) objects at low resolutions.
The popIII SFRD are weakly affected and do not suffer dramatically of resolution issues, although the corresponding contribution to the cosmic star formation reflects the limits mentioned about the total SFRD.
Such deviations are not crucial for our purposes, because they are not degenerate with the background dark-matter model, which predicts much larger effects at early times (see previous Figure~\ref{fig:sfr}).
If not explicitly noted -- as in the case of the following Figure~\ref{fig:wdmresolution} -- in the next sections we will mainly focus on the high-resolution runs sampled at $z=99$ with $512^3$ gas particles and $512^3$ dark-matter particles.

\begin{figure}
  \includegraphics[width=0.48\textwidth]{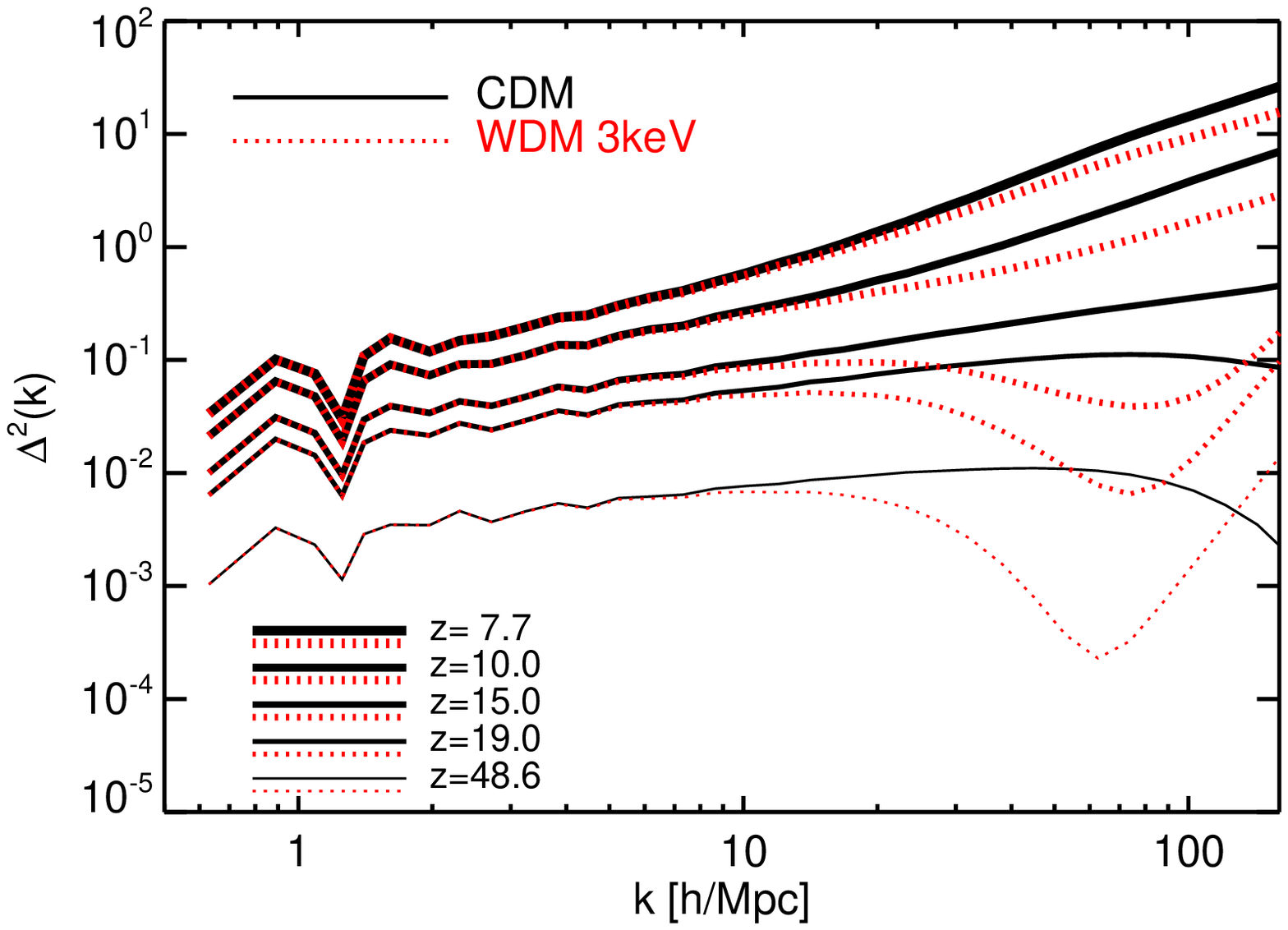}\\
  \vspace{-0.75cm}\\
  \includegraphics[width=0.48\textwidth]{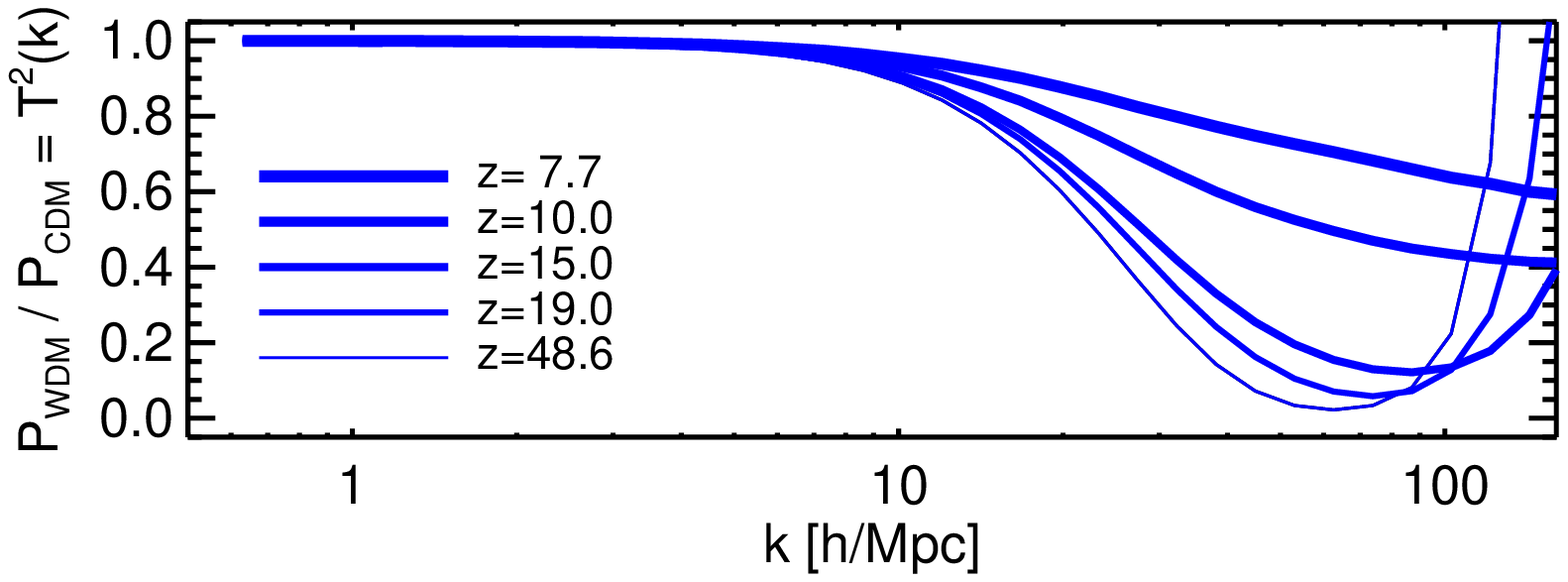}
  \caption[]{
    {\it Upper panel}.
    $\Delta^2$ computed from the non-linear matter power spectrum at different redshift (see legend). The black continuous curves refer to the CDM case, while the red dotted ones refer to the WDM case.
    {\it Lower panel}. Corresponding WDM transfer function squared, representing the ratio between WDM and CDM power spectra.
  }
  \label{fig:D2nl}
\vspace{-0.2cm}
\end{figure}

\subsection{Structure growth}\label{subsect:growth}

Non-linear evolution is responsible for birth and formation of small structures already from early epochs in both CDM and WDM, however, just the maps in the previous Figure~\ref{fig:maps} suggest that their clustering properties are not expected to evolve similarly.
In order to check this issue we compute the non-linear matter power spectra in the two scenarios at different redshifts and the corresponding mass functions.

The (dimensionless) CDM and WDM non-linear matter power spectra, $\Delta^2(k)$, are shown in the top panel of Figure~\ref{fig:D2nl}.
There we focus on the mode range $2\pi/L < k < k_{\rm Nyq}$, which is the physically significant one for our boxes of side $L$.
As time evolves, $\Delta^2$ evolves as well, growing by a few orders of magnitudes during the first billion years.
The power on larger scales (small $k$s) remain similar for both CDM and WDM models, while clustering from primordial smaller structures (large $k$s) induces trends in $\Delta^2$ that depend on dark-matter temperature.
This is due to the fact that smaller objects are more easily formed in CDM than in WDM (see also discussion later) and the resulting CDM non-linear power spectrum can grow faster at larger $k$.
In fact, for modes larger than $\sim 10\,h/{\rm Mpc}$ the discrepancy between CDM and WDM $\Delta^2$s reaches about two orders of magnitudes at $z\sim 50$ and roughly 1 dex at $z\sim 20$.
Only below $z\sim 15$ the two behaviours start to converge reaching a factor-of-a-few difference at $z \lesssim 10 $.
This is also consistent with the SFRDs previously discussed.
\\
In the bottom panel of Figure~\ref{fig:D2nl} we highlight the convergence of the clustering evolution by displaying the transfer functions squared at corresponding redshifts.
The epoch around $z\sim 15$ turns out to be quite interesting, because it reminds the onset of WDM formation, after which the WDM power spectrum catches up with the CDM one.
Effects of the shot noise in the WDM spectra, which can boost WDM power even above the CDM one at very large $k$, have negligible impacts on structure formation, as demonstrated in the following.

\begin{figure*}
\includegraphics[width=0.9\textwidth]{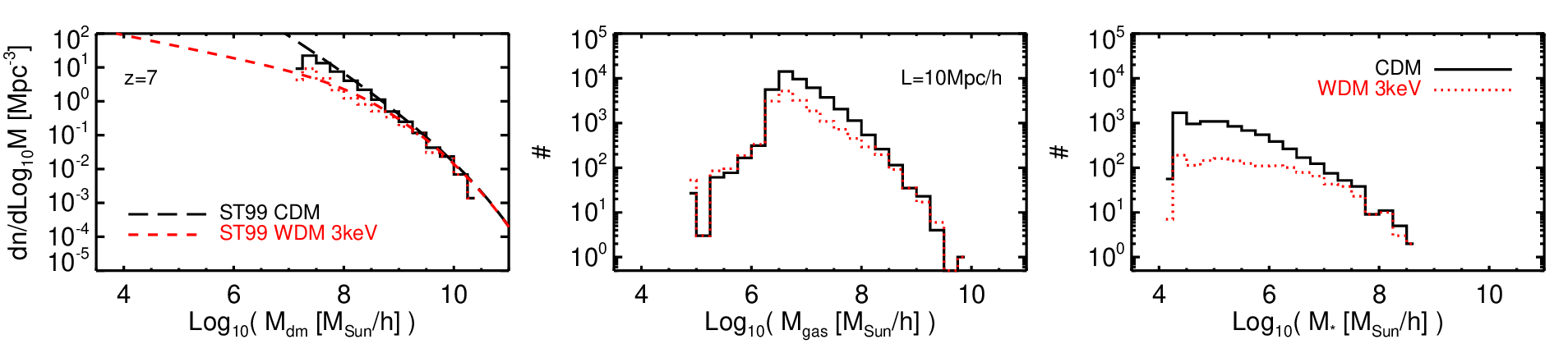}
\includegraphics[width=0.9\textwidth]{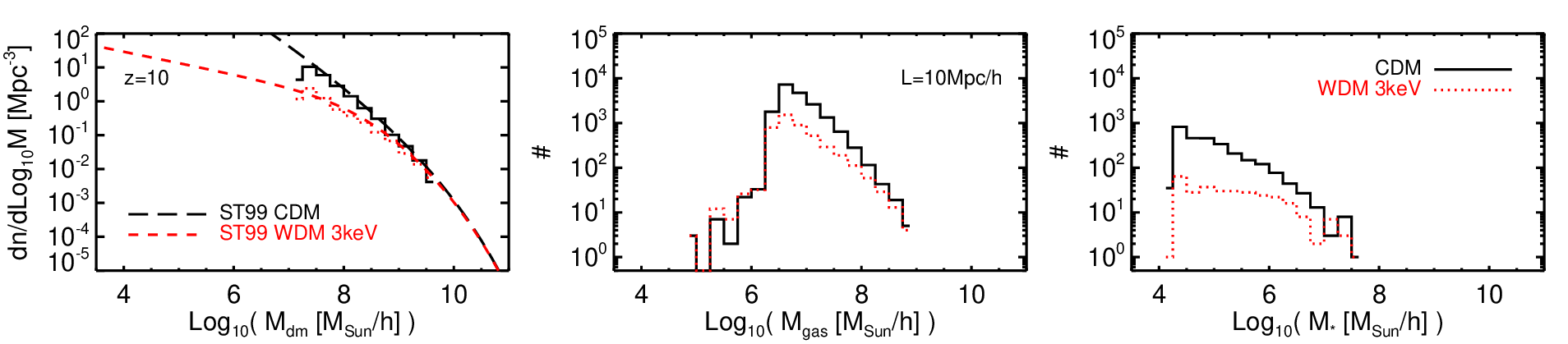}
\includegraphics[width=0.9\textwidth]{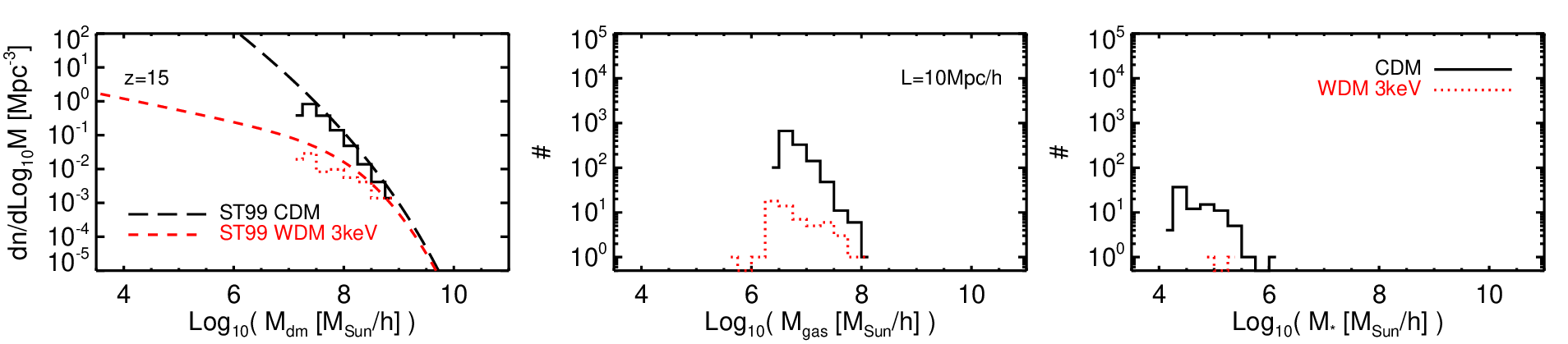}
\caption[]{
Dark (left), gaseous (center) and stellar (right) mass distributions in boxes of $\rm 10\, Mpc/{\it h}$ a side, both for the CDM (solid histograms) and for the WDM (dotted histograms) scenarios at redshift $z =$ 7, 10 and 15. Theoretical mass functions are computed by using \cite{ST1999} formalism (ST99) with CDM (long-dashed lines) and WDM (short-dashed lines) input power spectra, respectively.
}
\label{fig:halomasses}
\end{figure*}
In order to investigate more in depth the features of CDM and WDM structures we show in Figure~\ref{fig:halomasses} typical dark, gaseous and stellar masses of the cosmic objects at redshift $z=7$, $z=10$ and $z=15$, i.e. when the Universe was $\sim 1\,\rm Gyr$, $0.5\,\rm Gyr$ and $0.3\,\rm Gyr$.
In all the cases there are obvious differences in the abundance of CDM and WDM structures with smaller WDM objects being less common than the CDM counterparts.
Indeed, WDM haloes with masses $\lesssim 10^8\,\rm M_\odot/{\it h}$ are more than one order of magnitude rarer than CDM ones at $z=15$.
Such discrepancies get mitigated at later time, since at $z=10$ the difference is only a factor of $\sim 3-4$ and at $z=7$ reaches a factor of $\sim 2$.
The underlying reason for this is the cut in the WDM power spectrum at smaller scales, which causes a drop of the forming structures at any epoch.
Larger halo masses feature minor differences since they originate from regions of the power spectrum where CDM and WDM are virtually indistinguishable.\\
As a further check, we also compare the numerical dark-matter mass functions with the theoretical ones \cite[according to the classical formula by][labeled as ST99]{ST1999}.
The only difference between the CDM and WDM mass functions consists in the input power spectra (variance) employed in the calculations (see Figure~\ref{fig:Delta2}).
We find a reasonable agreement at any redshift, although $z=15$ samples are limited in statistics, mostly for the WDM case\footnote{
The number of WDM objects per each mass bin at $z=15$ is $\sim 1 - 10$.
}.
As already mentioned, even at the smallest scale sampled here we are not affected by numerical fragmentation for WDM structures, consistently with recent calculations by e.g. \cite{Schultz2014}.
Therefore, such issue seems not to be relevant for the first-Gyr objects.
\\
Gas masses experience similar statistical differences, since baryon evolution is generally hosted in dark-matter potential wells.
In fact, while the typical mass ranges are comparable at any $z$, the number of gas clouds for each mass bin in the WDM scenario is smaller than in the CDM one by a factor ranging from nearly $\sim 2$ dex at $z=15$ down to a few at $z=7-10$.\\
Similarly, WDM stellar distributions are under-abundant with respect to CDM at all times.
Discrepancies in stellar abundances remain as high as one dex even after $\sim 1\,\rm Gyr$ and experience a delayed catch-up due to the typically low (a few per cent) star formation efficiencies and gas-to-star conversion factors \cite[e.g.][]{BiffiMaio2013}.
Shortly, the lack of gas clouds and proto-galaxies in the WDM universe is essentially shaped by the corresponding lack of star formation sites hosted by dark-matter haloes (see next discussion).

\begin{figure}
\centering
\includegraphics[width=0.45\textwidth]{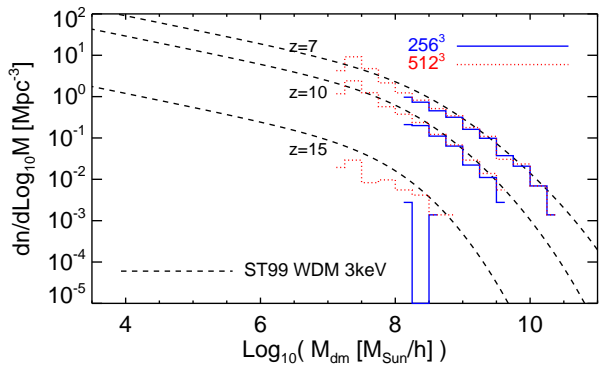}
\caption[]{
Dark-matter mass functions for two WDM simulations with resolution of $256^3$ and $512^3$ in boxes with $10\,\rm Mpc/{\it h}$ side at redshift $z =$ 7, 10 and 15.
Theoretical mass functions are computed according to the \cite{ST1999} formalism.
}
\label{fig:wdmresolution}
\end{figure}

We test our results against resolution issues in Figure~\ref{fig:wdmresolution} by considering the low-resolution WDM run ($256^3$ particles for the dark species). The figure displays the mass functions at different redshift for both the $256^3$ and the $512^3$ set-ups.
In all the cases we see converging trends in the mass ranges sampled by the different resolutions.
Same completeness problem arises for the $256^3$ run at $z=15$ due to the inability of such set-up to capture the birth of small primordial objects that lie below the resolution limit.
The behaviour of the simulated mass functions is in line with theoretical expectations, as well, and does not present hints of spurious clustering due to shot noise (see figure \ref{fig:D2nl}), as argued in Sect.~\ref{sect:runs}.

\subsection{Star formation activity}\label{subsect:SFactivity}

\begin{figure}
\includegraphics[width=0.48\textwidth]{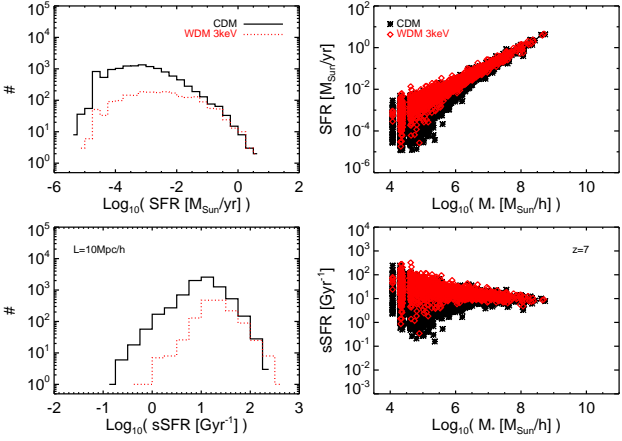}
\caption[]{SFR (top left) and sSFR (bottom left) distributions for structures in the CDM and in the WDM scenarios at $z=7$ with corresponding $ M_\star - \rm SFR$ (top right) and $ M_\star - \rm sSFR$ (bottom right) relations.
}
\label{fig:sSFR7}
\end{figure}

Figures~\ref{fig:sSFR7} displays the basic properties of star forming structures at redshift $z=7$.
SFR and specific SFR (sSFR) distributions are reported on the left panels, while their dependence with mass is shown on the right panels.
Typical SFRs range between $\sim 10^{-5}$ and $\sim 10 \,\rm M_\odot/yr$, the distributions are quite broad both in WDM and in CDM and the fractions of objects that have formed stars by $z=7$ converges to similar values ($70\%$ and $67\%$, respectively).
The number of proto-galaxies with $\rm SFR \lesssim 10^{-2}\,\rm M_\odot/yr$ is about 1 dex larger in the CDM models, though.
Resulting sSFRs are around $\sim 10^{-1} - 10^2 \,\rm Gyr^{-1} $ with a clear peak at $\sim 10\,\rm Gyr^{-1}$, consistently with expected values at those epochs \cite[e.g.][]{Salvaterra2013, Maio2013dla}.
Due to the inhibited collapse of smaller masses, WDM predictions show a sensitive paucity of objects with $\rm SFR\lesssim 10^{-2}\,M_\odot/yr$ and a complete lack of the quiescent population with sSFR below $\sim 1\,\rm Gyr^{-1}$, whereas CDM objects in the same range are still common.
This is in line with our discussion in previous section~\ref{subsec:evolution} and implies that WDM structures, albeit rarer, host typically stellar populations whose onset is more sudden, or bursty, than CDM (as it will be also supported by the fraction of star hosting haloes discussed below).
The statistical differences have little impacts on the correlation between stellar mass and star formation activity, as demonstrated by the $M_\star - \rm SFR$ and $ M_\star - \rm sSFR$ relations in the right panels of the figure.
Both turn to be quite independent from the dark-matter nature and seem mainly led by the baryon processes occurring within the structure potential wells.
Consequently, SFR scales almost linearly with $M_\star$ and sSFR features a weak decrease with $M_\star$ in both WDM and CDM.

\begin{figure}
\includegraphics[width=0.48\textwidth]{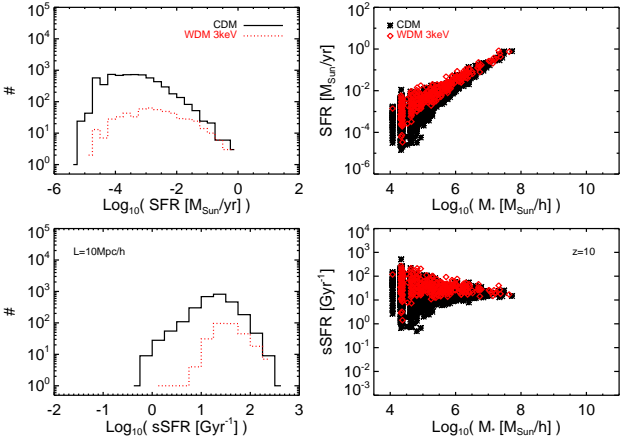}
\caption[]{
  Same as Figure~\ref{fig:sSFR7} at $z=10$.
}
\label{fig:sSFR10}
\end{figure}

\begin{figure}
\includegraphics[width=0.48\textwidth]{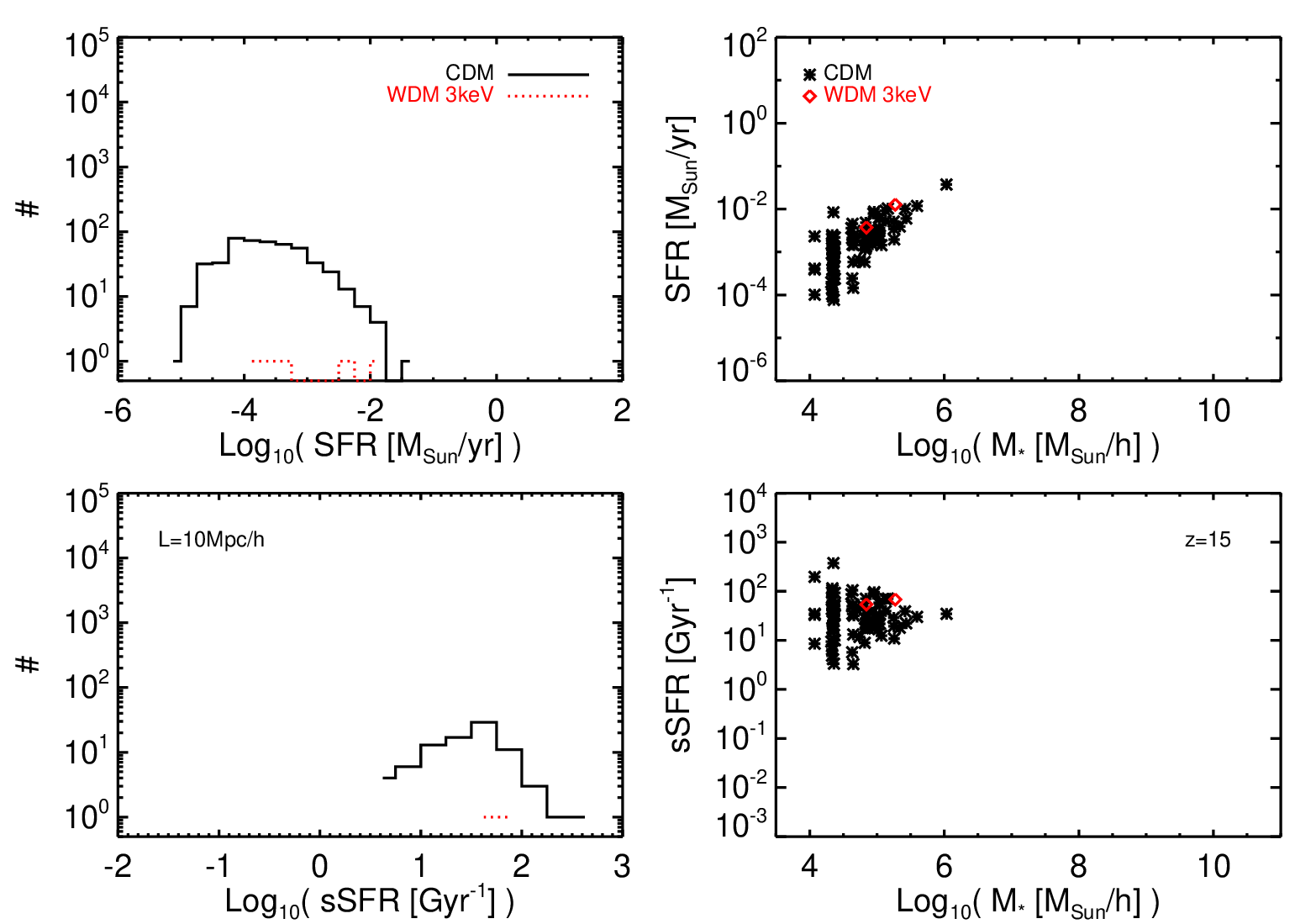}
\caption[]{
  Same as Figure~\ref{fig:sSFR7} at $z=15$.
}
\label{fig:sSFR15}
\end{figure}

Similar considerations hold for the $z=10$ samples, displayed in Figure~\ref{fig:sSFR10}.
At this time, SFRs are below $\sim 1\,\rm M_\odot/yr$, with typical values between $\sim 10^{-4}$ and $10^{-2}\,\rm M_\odot/yr$, while sSFRs are larger than $\sim 1\,\rm Gyr^{-1}$ in CDM and larger than $\sim 10\,\rm Gyr^{-1}$ in WDM, with peak values of $\sim 20-30 \,\rm Gyr^{-1}$, respectively.
SFRs and sSFRs still lie within the expected trends for $M_\star - \rm SFR$ and $ M_\star - \rm sSFR$ relations.
The CDM results are in agreement with previous studies of primordial galaxies \cite[e.g.][]{BiffiMaio2013, Wise2014}, while for the WDM case they are consistent with the general behaviour discussed above.
We highlight the powerful star formation activity in the (rarer) WDM haloes, whose majority has $\rm sSFR > 10\, Gyr^{-1}$.
This represents a neat evidence of the sudden star formation activity established in those WDM structures which have finally grown up to scales with sufficient power to collapse.
In CDM haloes, the continuously decreasing power-law trend of the power spectrum at large $k$ determines a smoother growth and a milder star formation onset (as seen in Figure~\ref{fig:sfr}).
The overall consequence is that the fraction of CDM haloes that formed stars by $z=10$ is $43\%$ while for WDM haloes this fraction is as high as $55\%$: i.e., in the scenario with WDM there are less structures, but they form stars more actively.
The situation is even more evident when looking at the very initial stages of baryon assembly at $z=15$ (Figure~\ref{fig:sSFR15}).
The few WDM objects (5) undergoing star formation have sSFR$\sim 10^2\,\rm Gyr^{-1}$ and $40\%$ of them (2) have already formed stars.
On the other side, CDM objects have already well shaped SFR and sSFR distributions, sampled with about 500 star forming proto-galaxies.
Their peak values are at $\rm SFR\lesssim 10^{-4}\,\rm M_\odot/yr$ and at $ \rm sSFR \sim 10^{1.5}\,\rm Gyr^{-1}$, respectively.
However, only $17\%$ of the CDM hosting haloes have formed stars.
The resulting $M_\star$ shows some scaling with SFR and sSFR in the CDM universe, while it is not possible to draw firm conclusions for the WDM case, due to poor statistics.

For a better comparison, the fraction of haloes which host stars is reported for both dark-matter models at different redshifts in Table~\ref{table:SFhost}.
This latter highlights the converging trend at the end of the first billion years and the evident differences at the beginning of the cosmic dawn.
\\
We note that, despite the initial delay in the formation of proto-galaxies in WDM haloes, the primordial Universe should have been very bursty and, hence, should have produced observationally detectable sources in both CDM and WDM scenarios.
\\
From a statistical point of view, the amount of proto-galaxies formed in a CDM cosmology is up to 100 times larger than WDM and, in the next future, this could help pose serious constraints on the nature of dark matter by employing extensive searches of luminous objects at redshift $z > 10$.
\\
Observationally (see more detailed discussion in the following Sect.~\ref{subsect:LF}), the significant lack of WDM structures during the dark ages might have repercussions not only on the formation of the lowest-mass haloes (e.g. dwarf galaxies), but also on the consequent onset of reionization and cosmic heating (spin temperature) of neutral hydrogen, due to the sparser and retarded evolution of HI brightness temperatures expected in a WDM universe.
\begin{table}
  \caption{Fraction of star hosting haloes in CDM and WDM models.}
  \centering
  \begin{tabular}{l c c}
    \hline
    \hline
    Redshift \phantom{xxx} & \phantom{xxx} CDM \phantom{xxx} &  \phantom{xxx} WDM  \phantom{xxx} \\
    [0.5ex]
    \hline
    $z=7$   &  67~\%  &  70~\% \\
    $z=10$  &  43~\%  &  55~\% \\
    $z=15$  &  17~\%  &  40~\% \\
    [1ex]
    \hline
  \end{tabular}
  \label{table:SFhost}
\end{table}

\subsection{Baryon fractions}\label{subsect:fractions}
Star formation and feedback mechanisms can have further implications on the baryon evolution in different haloes.
Since the cosmic star formation activity depends on the background model, we might expect some implications for gas ($f_{\rm gas}$) and stellar ($f_{\star}$) fractions at different times.

\begin{figure}
\includegraphics[width=0.48\textwidth]{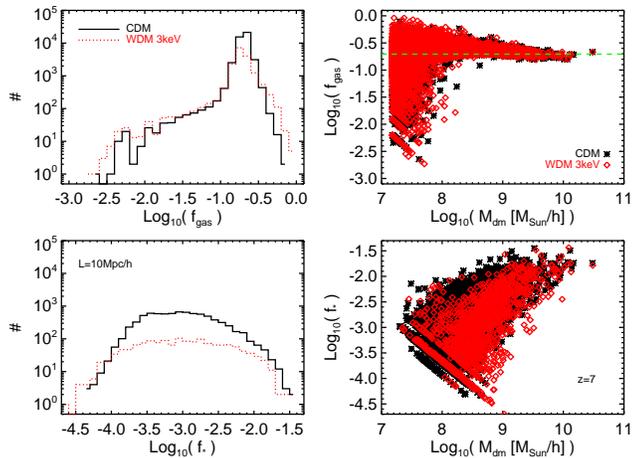}
\caption[]{
  Gas (top left) and stellar (bottom left) fraction distributions at $z=7$ and corresponding $f_{\rm gas} - M_{\rm dm}$ (top right) and $f_{*} - M_{\rm dm}$ (bottom right) relations, both in the CDM and in the WDM (3 keV) scenarios. The horizontal dashed line corresponds to the baryon fraction expected by the given input cosmological parameters.
}
\label{fig:fractions7}
\end{figure}

Figure~\ref{fig:fractions7} displays $f_{\rm gas} $ and $f_{\star}$ distributions at $z=7$ (left panels) and corresponding dependencies on the dark mass of the hosting haloes (right panels).
The gas behaviour highlights a peak in correspondence of the cosmic baryon fraction and a large spread of about 3 orders of magnitudes due to the ongoing stellar-feedback mechanisms that heat up the gas and evacuate material from smaller structures, once the onset of star formation has taken place.
A neat evidence of feedback mechanisms is the spread in $f_{\rm gas}$ at low masses, caused by stellar heating and/or winds. In fact, these processes can evacuate material in nearby satellite haloes or replenish other haloes with additional gas. Such phenomena are particularly evident in very small haloes, because their potential wells are not deep enough to trap the gas when this is accidentally heated by external sources.
Discrepancies between WDM and CDM $f_{\rm gas}$ distributions are within a factor of 2, because, as long as dark-matter haloes are in place, gas collisional properties result much more sensitive to local baryon physics than to the nature of the hosting dark-matter structure.
\begin{figure}
\vspace{-0.2cm}
\includegraphics[width=0.48\textwidth]{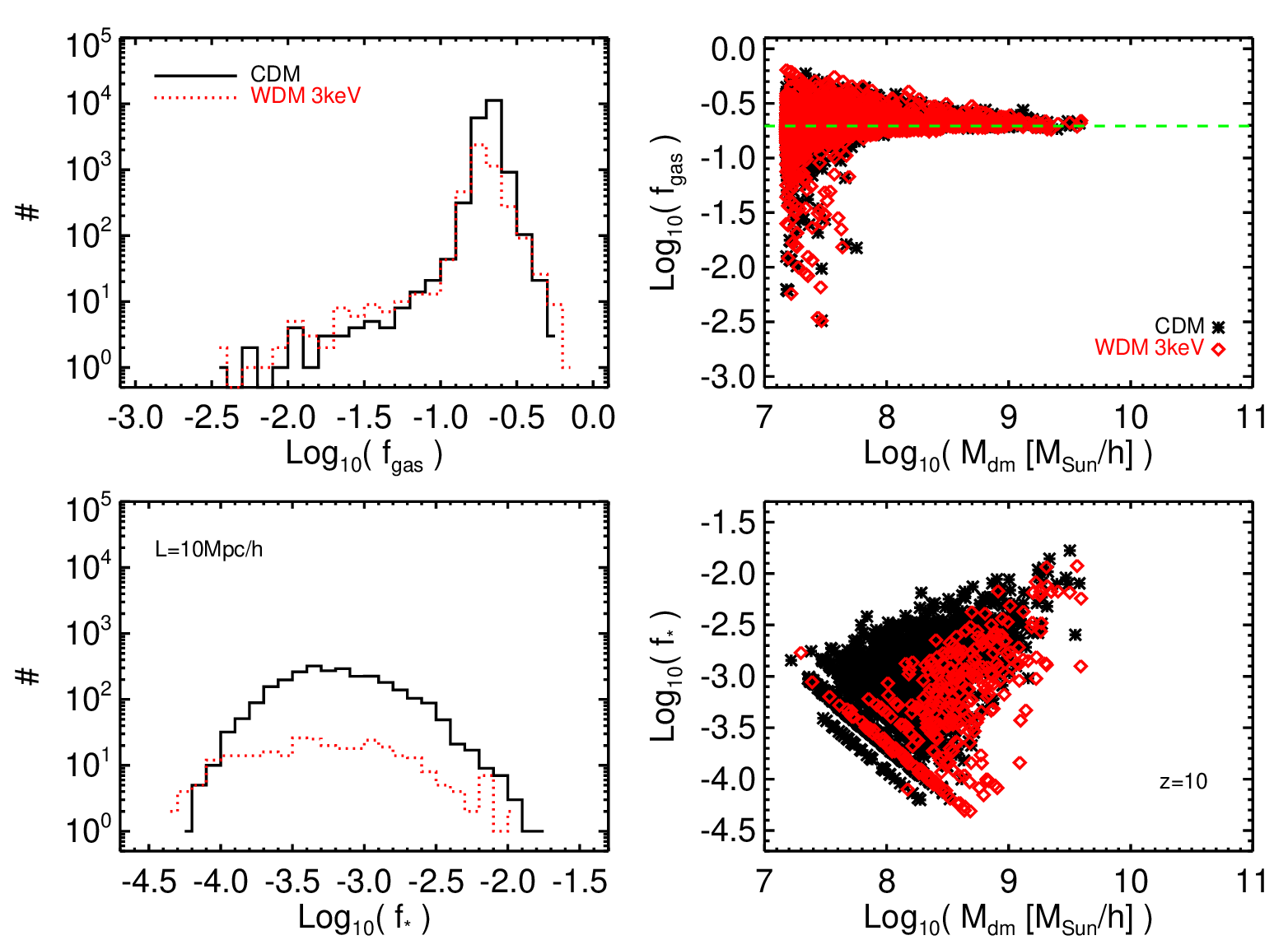}
\caption[]{
  Same as Figure~\ref{fig:fractions7} at $z=10$.
}
\label{fig:fractions10}
\vspace{-0.3cm}
\end{figure}
\begin{figure}
\includegraphics[width=0.48\textwidth]{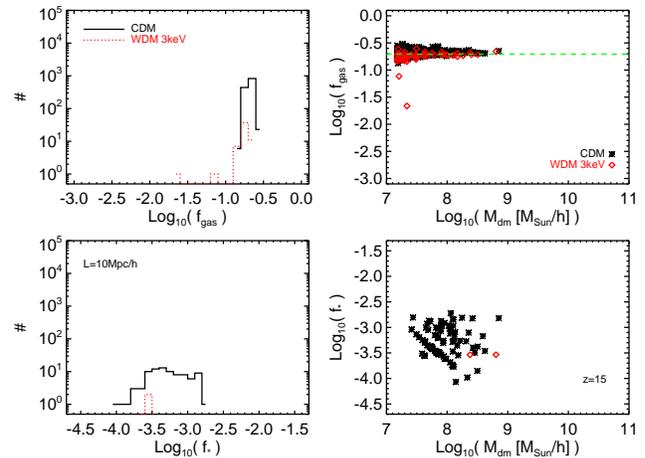}
\caption[]{
  Same as Figure~\ref{fig:fractions7} at $z=15$.
}
\label{fig:fractions15}
\end{figure}

The distributions of $f_\star$, however, are quite different. This is due to the tight dependence of the runaway molecular collapse and star formation rate on the adopted dark-matter scenario (see previous discussion).
Therefore, star production in CDM is enhanced with respect to WDM up to a factor of 10 at $f_{\star}\sim 10^{-4}-10^{-2}$.
Despite the implications for their statistical distributions, gas and stellar fractions exhibit similar trends with $M_{\rm dm}$ in both CDM and WDM (right panels).
The shape of $f_{\rm gas}$ as a function of mass is led at high $M_{\rm dm}$ by the gravitational potential of larger haloes, that can trap easily gaseous material even though this has been heated by star formation feedback.
At lower masses, instead, halo potential wells are not deep enough to retain the material evacuated by star formation feedback (either in the same halo or in nearby haloes) and the corresponding gas content can drop by a few orders of magnitude.
This argument holds both for CDM and for WDM models and can be considered as an evidence that the features of baryonic matter within star forming haloes are more strongly dominated by their environmental chemical and thermodynamical properties rather than their dark properties.
On the other hand, $ f_\star $ increases as a function of dark-matter halo mass in both CDM and WDM, with a steeper increase for WDM, following the star formation rate.

Results going in the same direction are found also at all the other redshifts (see e.g. Figure~\ref{fig:fractions10} and Figure~\ref{fig:fractions15}) with the peculiarity that at higher $z$ the cascade effects of the cut in the WDM power spectrum become more evident.
In fact, at $z=10$ the $f_{\star}$ distribution in the WDM model is more than 1 dex lower than in the CDM model, while at $z=15$ there are only a few WDM haloes that host stars.
Gas fractions are still close to the cosmic fraction, because feedback effects have not yet had time to act significantly.
This confirms that the statistical differences in the two dark-matter scenarios are essentially due to the fact that those scales that could collapse and form stars in the CDM model at $z \gtrsim 15$ are still inactive in the model with WDM.

\subsection{Molecules and metals in the early objects }\label{subsect:chemistry}

\begin{figure*}
\centering
\includegraphics[width=0.33\textwidth]{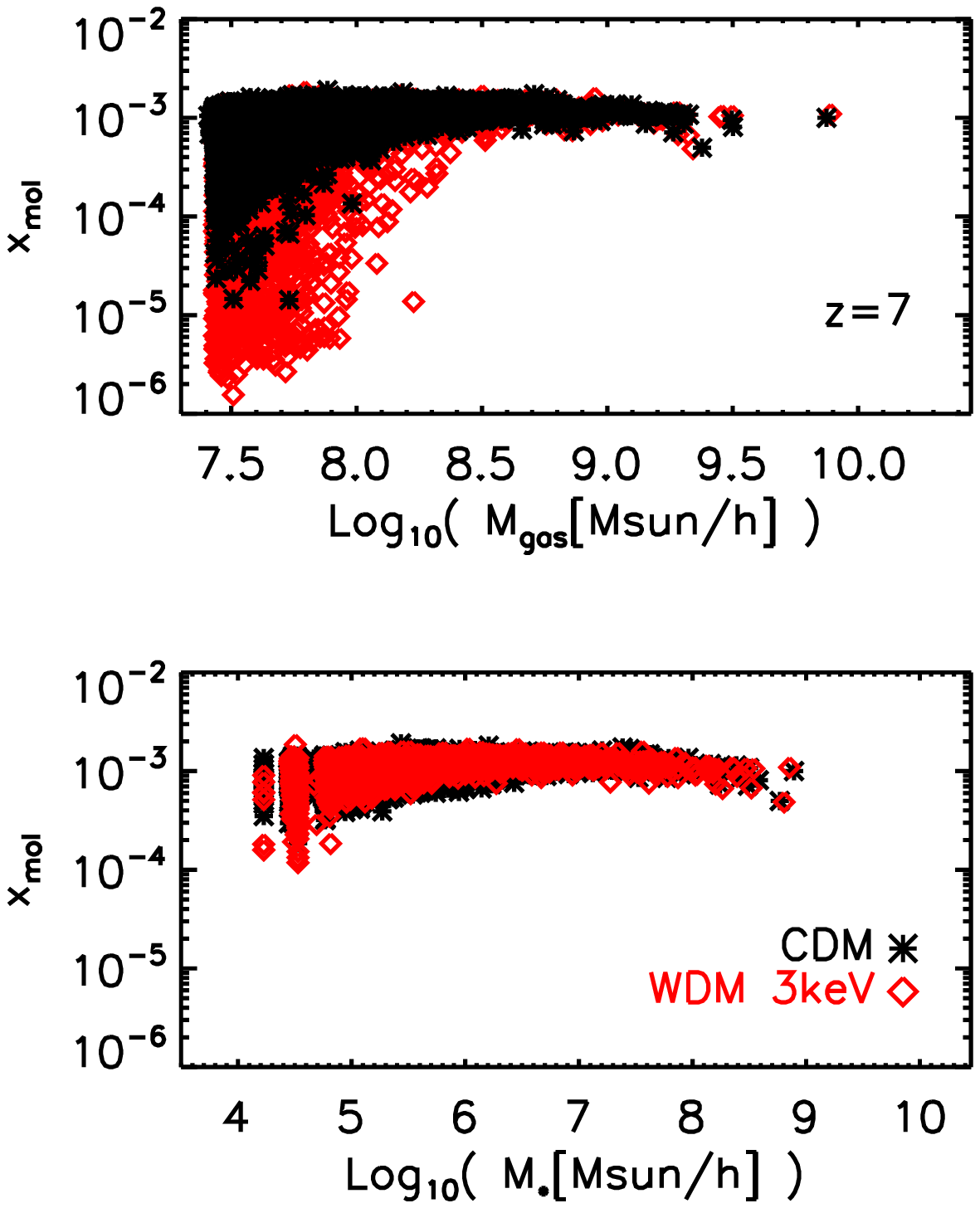}
\includegraphics[width=0.33\textwidth]{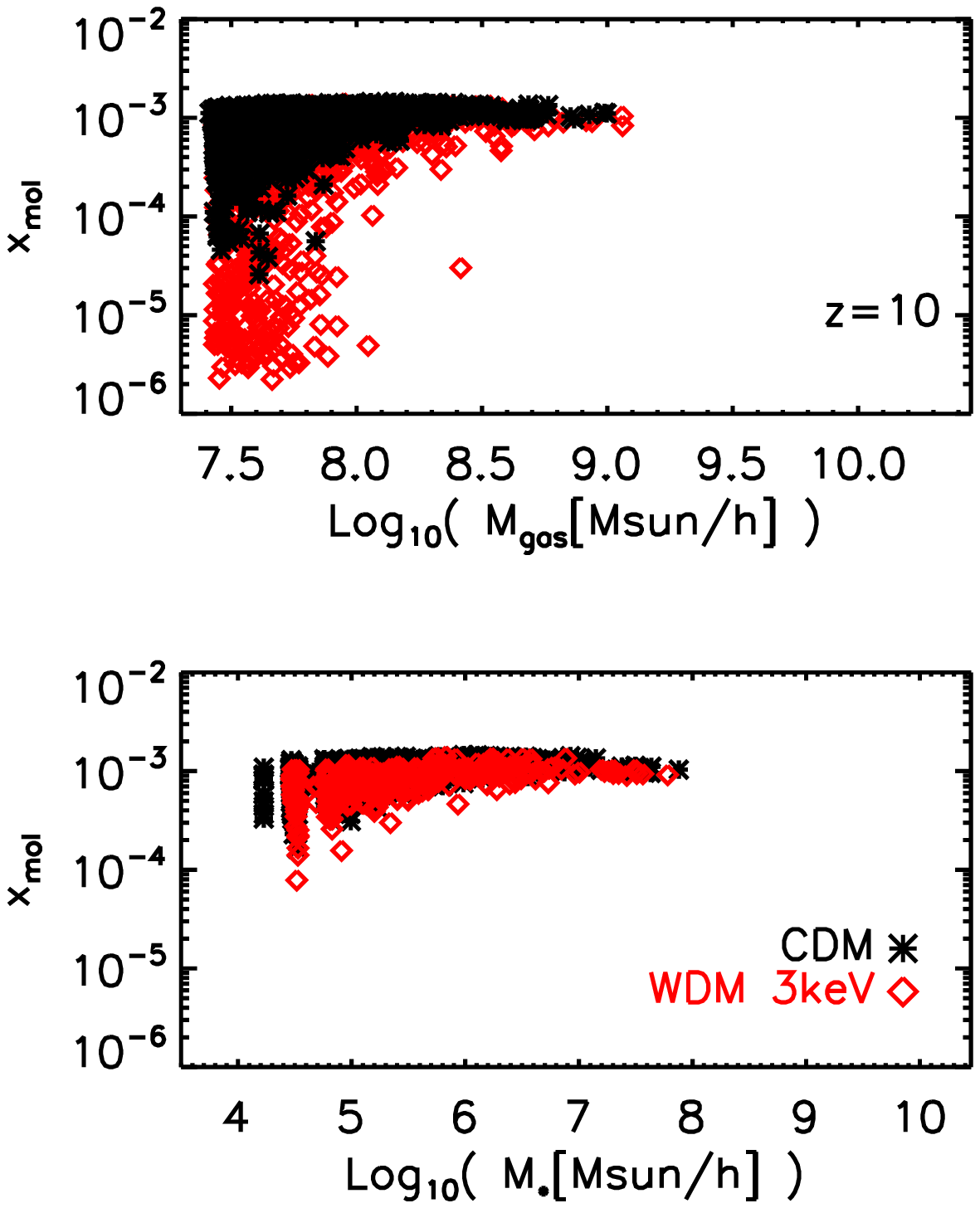}
\includegraphics[width=0.33\textwidth]{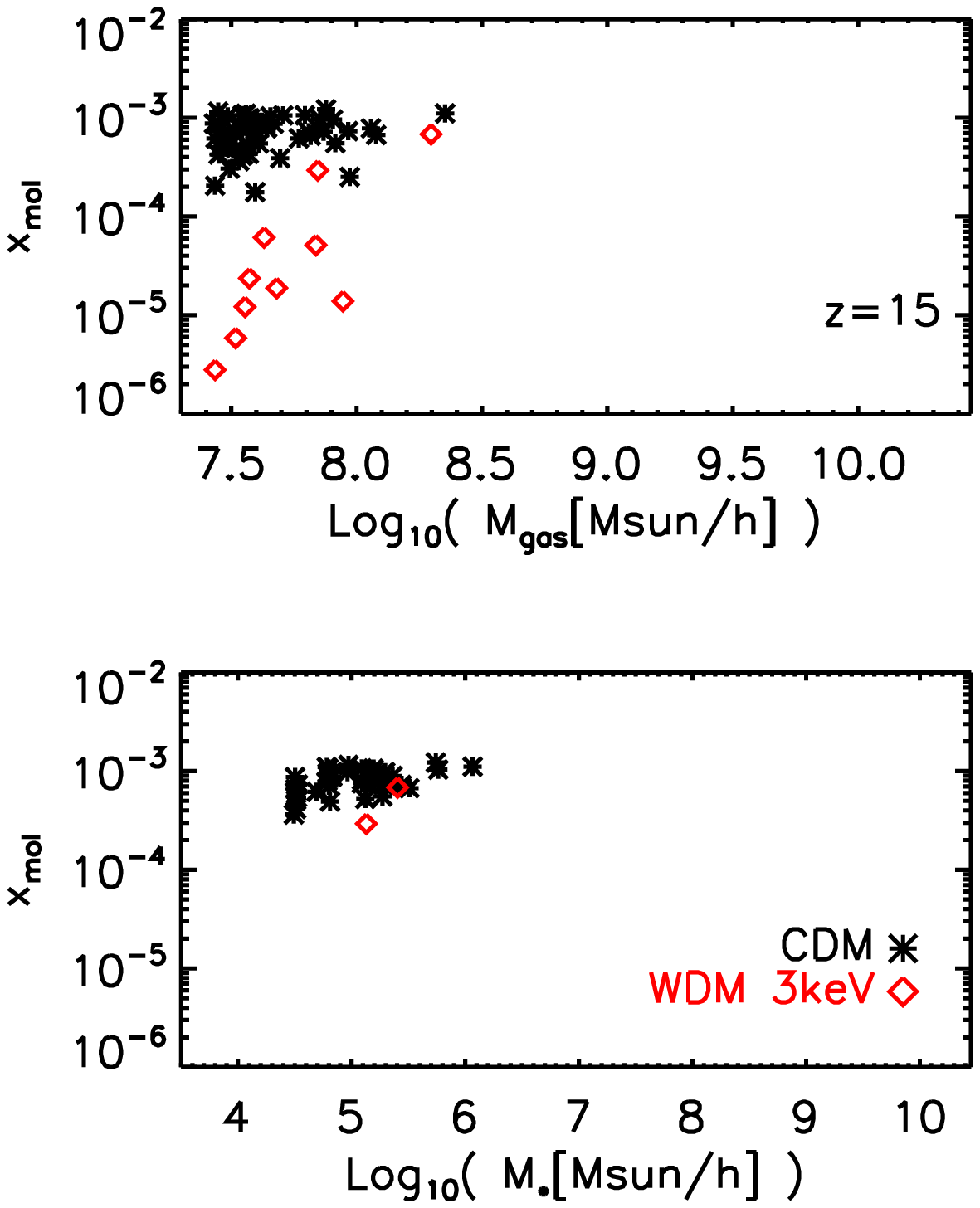}
\caption[]{
Molecular content, $x_{\rm mol}$, as a function of gas mass (upper panels) and stellar mass (lower panels) at redshift $z=7$ (left), $z=10$ (center) and $z=15$ (right) for both CDM (black asterisks) and WDM (red empty squares).
}
\label{fig:xmol}
\end{figure*}
 In Figure~\ref{fig:xmol} we investigate the trends of the molecular content as a function of gas (upper panels) and stellar (lower panels) mass at redshift $z=7$, 10 and 15.
Molecular content is quantified by computing the molecular fraction, $x_{\rm mol}$, as the ratio between the H$_2$ mass and the gas mass of each cosmic structure. For the sake of simplicity, we focus here only on those structures whose molecular content can be safely sampled by at least 300 gas particles and corresponding roughly to $\sim 20$ per cent of the whole sample. This choice allows us to rapidly compare structures that have similar growth. The scatter plots in figure demonstrate the evolution of molecular content that is built up over a cosmic time of a fraction of Gyr.
\\
At $z=15$ CDM objects with gas masses around $\sim 10^8\, \rm M_\odot$ (the tail of the mass distribution at these epochs) tend to have typical $x_{\rm mol}\sim 10^{-4}-10^{-3}$.
These (proto-)galaxies are able to form stars at a rate of up to $\sim 10^{-2}-10^{-1}\,\rm M_\odot/yr$, as shown before, and have already converted up to $\sim  0.5$ per cent of their gas into stars.
As mentioned, WDM objects suffer a general delay which translates into a lower number of structures formed at any given time.
Their molecular content is rapidly catching up with respect the CDM behaviour and $x_{\rm mol}$ ranges between $\sim 10^{-6} -10^{-3} $.
Two objects have $x_{\rm mol}\sim  10^{-4} -10^{-3} $, resulting $\rm SFR \sim 10^{-2}\,\rm M_\odot/yr$ and gas-to-star conversion efficiency of $\sim 0.1$ per cent.
The other WDM structures are still in a state of gaseous clumps (note the lack of points in the lower panel) that will form stars only in subsequent~phases.
\\
Within about half Gyr, more structures form and host gas collapse through runaway cooling, but, despite the statistical improvements, the same picture outlined before is recovered.
At $z=10$ one can still see that typical $x_{\rm mol}$ in CDM galaxies are higher and less scattered than WDM galaxies and the region at $x_{\rm mol} < 10^{-4}$ is predominantly populated by gaseous structures in WDM haloes.
SFRs reach $\gtrsim 1\,\rm M_\odot/yr$ and resulting stellar masses have increased up to $\sim 10^8\,\rm M_\odot/yr$.
Gas-to-star conversion efficiencies are as high as $\sim 10$ per cent, possibly enhanced by cooling due to the metals expelled by the first SNe.
\\
Similar trends are found also at roughly 1 Gyr of cosmic time, $z=7$.
At this epoch both CDM and WDM universe are quite active, with $\rm SFR \gtrsim 10\,\rm M_\odot/yr$ and stellar feedback becoming more and more important in regulating star formation processes.
The implications of ongoing feedback effects can be recognised in the $x_{\rm mol}$ patterns of both WDM and CDM, with an increasingly larger number of structures featuring $x_{\rm mol}\sim 10^{-5} - 10^{-4} $, as a consequence of thermal dissociation of H$_2$.
These objects do not form stars, though (lower panel), consistently with $z=10$ and $z=15$ panels, and are affected by star formation events in their neighbourhood.
\begin{figure}
\centering
\includegraphics[width=0.45\textwidth]{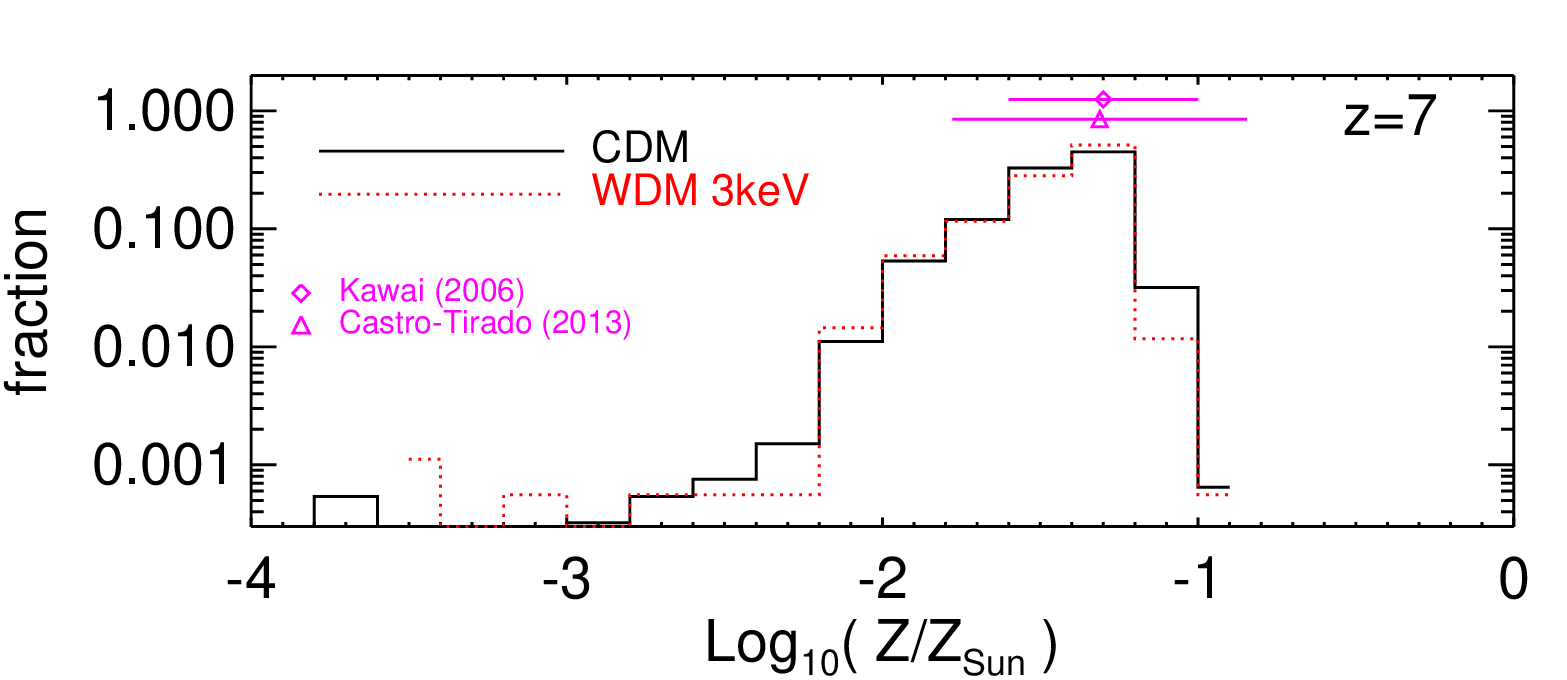}
\includegraphics[width=0.45\textwidth]{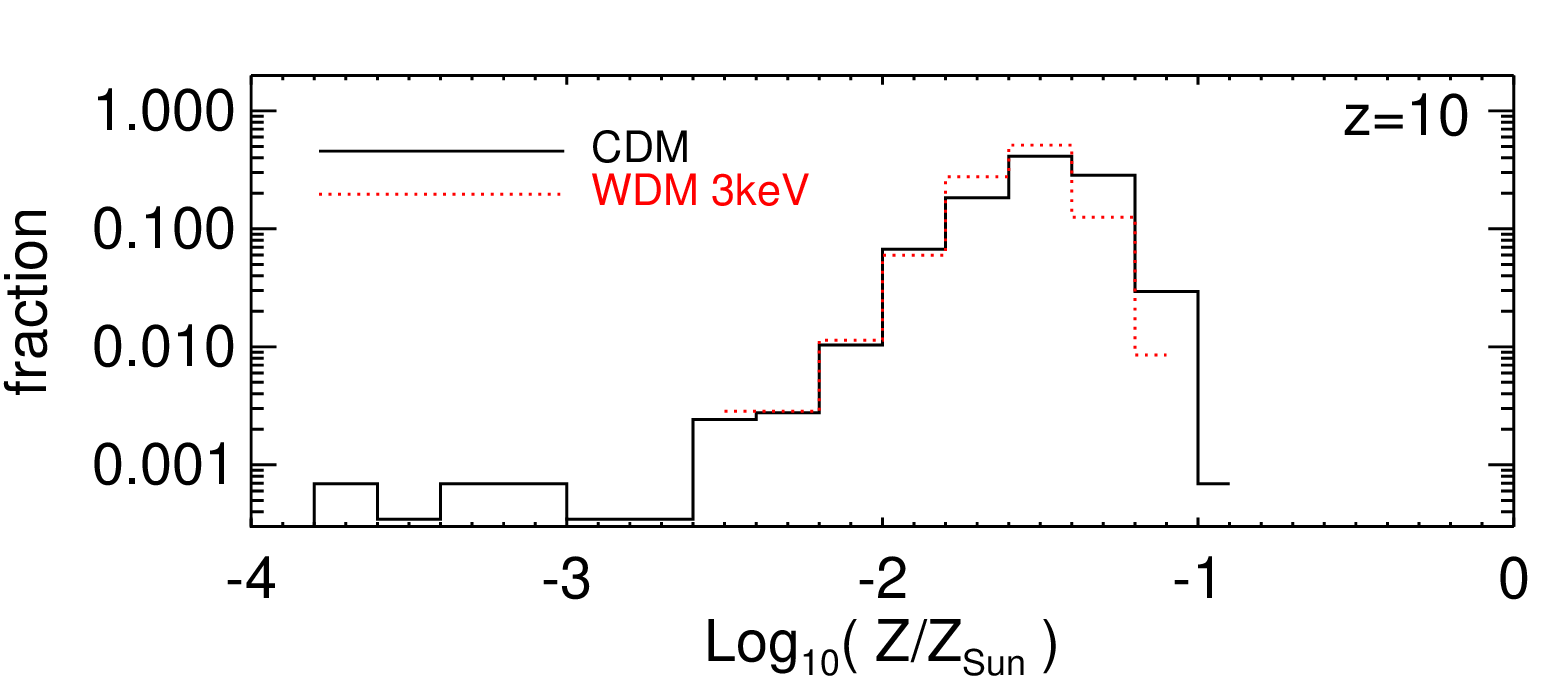}
\caption[]{
  Normalized metallicity distributions for CDM (black solid line) and WDM (red dotted line) at redshift $z=7$ (top) and $z=10$ (bottom) with observational determinations derived by long-GRB host galaxies at redshift $z\simeq 6.3$ by \cite{Kawai2006} and at $z\simeq 5.9 $ by \cite{Castro2013}.
}
\label{fig:Zdistributions}
\end{figure}

As we said, heavy elements can give a relevant contribution to the determination of the dominant stellar population and to the cooling capabilities of the gas in bound structures.
Therefore, it is interesting to check whether metallicity distribution might result affected in the two different dark-matter models.
\\
In Figure~\ref{fig:Zdistributions} we plot the $Z$ distributions at $z=10$ and $z=7$ for both CDM and WDM.
We neglect the $z=15$ due to the paucity of WDM objects.
Since the two models have a significantly different number of structures (see mass functions in the initial discussion of the results), to more directly compare them  we normalized the distributions to unity in both cases.
The resulting $Z$ values range between $\sim 10^{-1}$ and $10^{-3}\,Z_\odot$ with only a couple of outliers at lower metallicities ($< 0.1$ per cent level).
Metallicity peaks for CDM and WDM are both around $10^{-1.5}\,Z_\odot$ and most of the data lie between $0.01$ and $0.1$.
Peaks for CDM and WDM at $z=7$ agree with the two available observational determinations of metallicity in high-$z$ long-GRB host galaxies and their distributions do not present significantly different features.
This is due to the fact that the dominant pollution events in the first Gyr are massive-SN explosions, which have extremely short lifetimes of $\sim 10^6 - 10^7 \,\rm yr$ and lead metal spreading in both scenarios.
Details in the implementation of winds (as long as wind velocities are in the range $\sim 300-500\,\rm km/s$) are not expected to influence strongly metallicity distributions at these epochs, because typical cosmic structures are still quite small and can easily experience gas evacuation. They could have implications at lower $z$, though \cite[][]{Tescari2014}.
In conclusion, it seems that the time delay of WDM growth with respect to CDM at $z\sim 15 -10$ is responsible for a later collapse and a smaller statistical sample, but this does not affect the resulting typical $Z$s that are determined only by stellar yields.
\begin{figure*}
\centering
\includegraphics[width=0.33\textwidth, height=5cm]{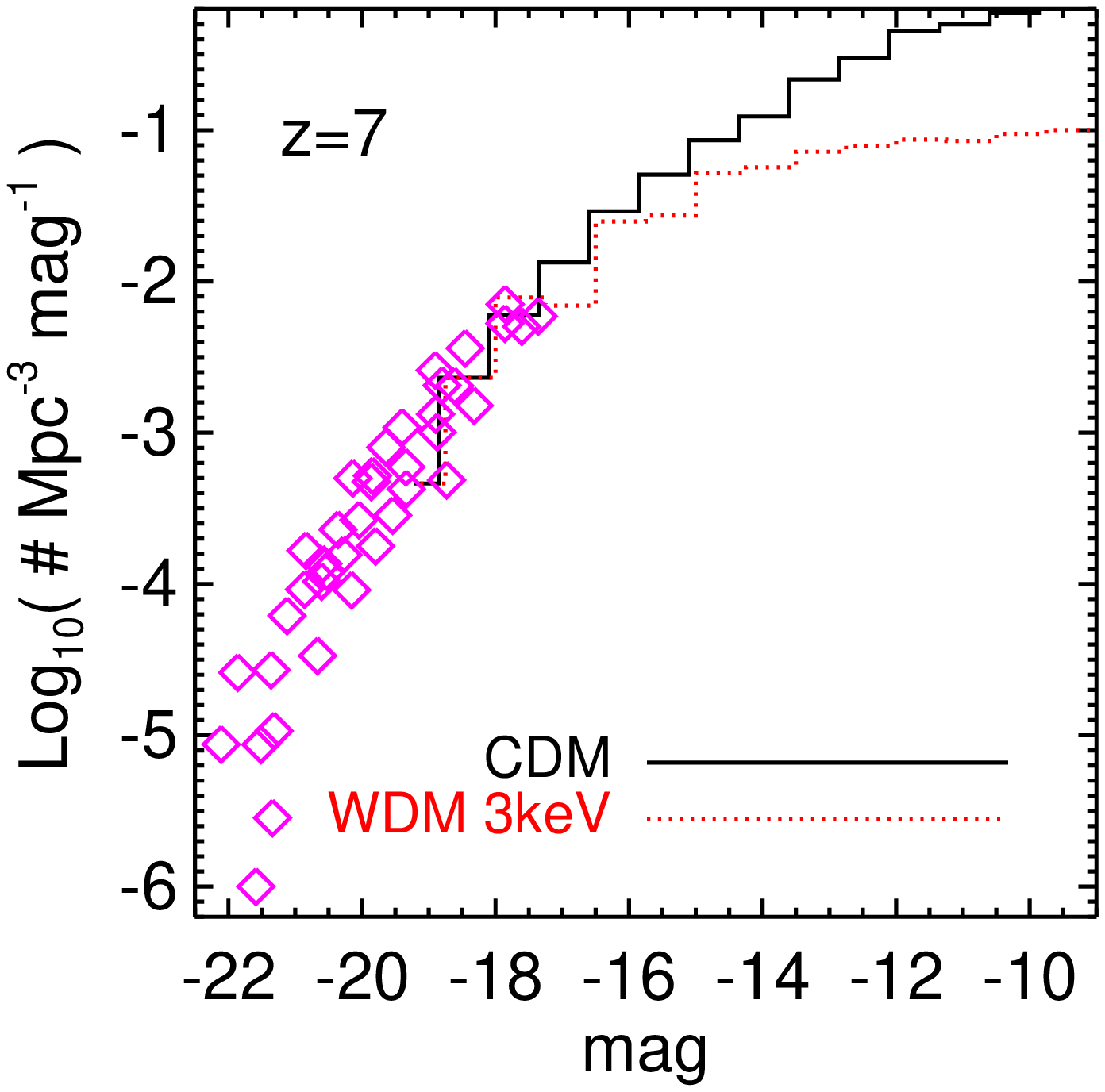}
\includegraphics[width=0.33\textwidth, height=5cm]{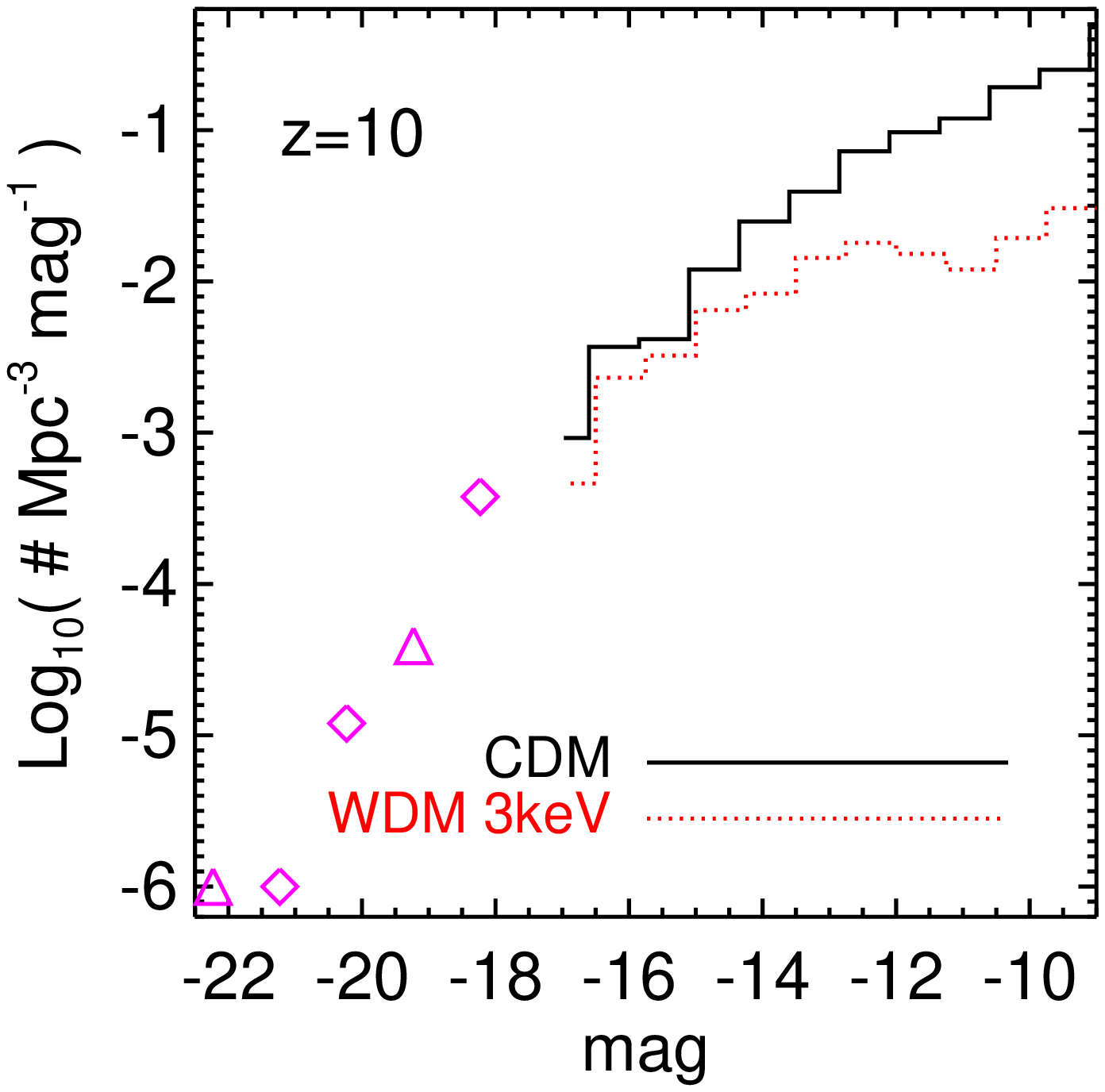}
\includegraphics[width=0.33\textwidth, height=5cm]{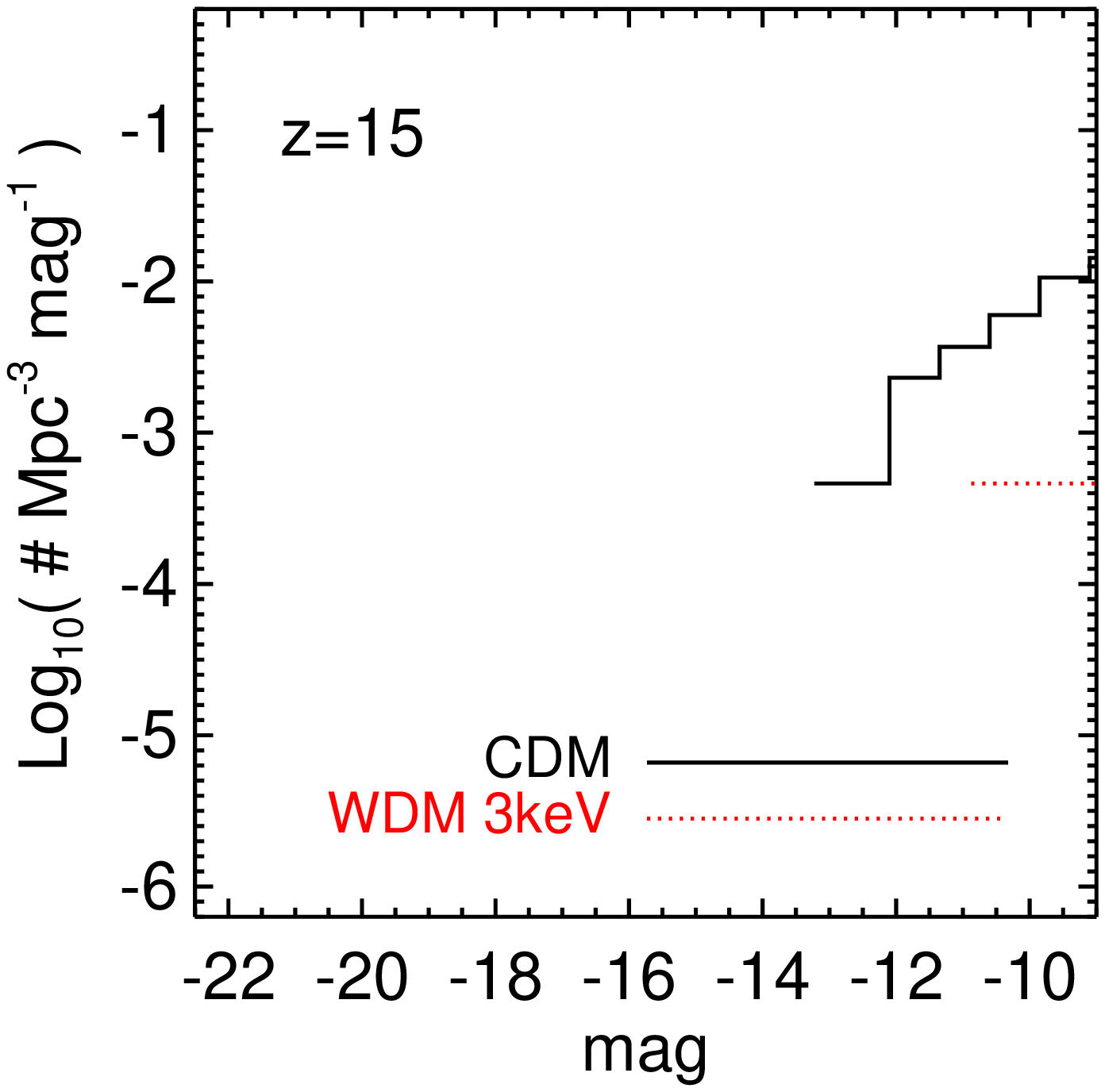}\\
\vspace{-0.1cm}
\includegraphics[width=0.33\textwidth]{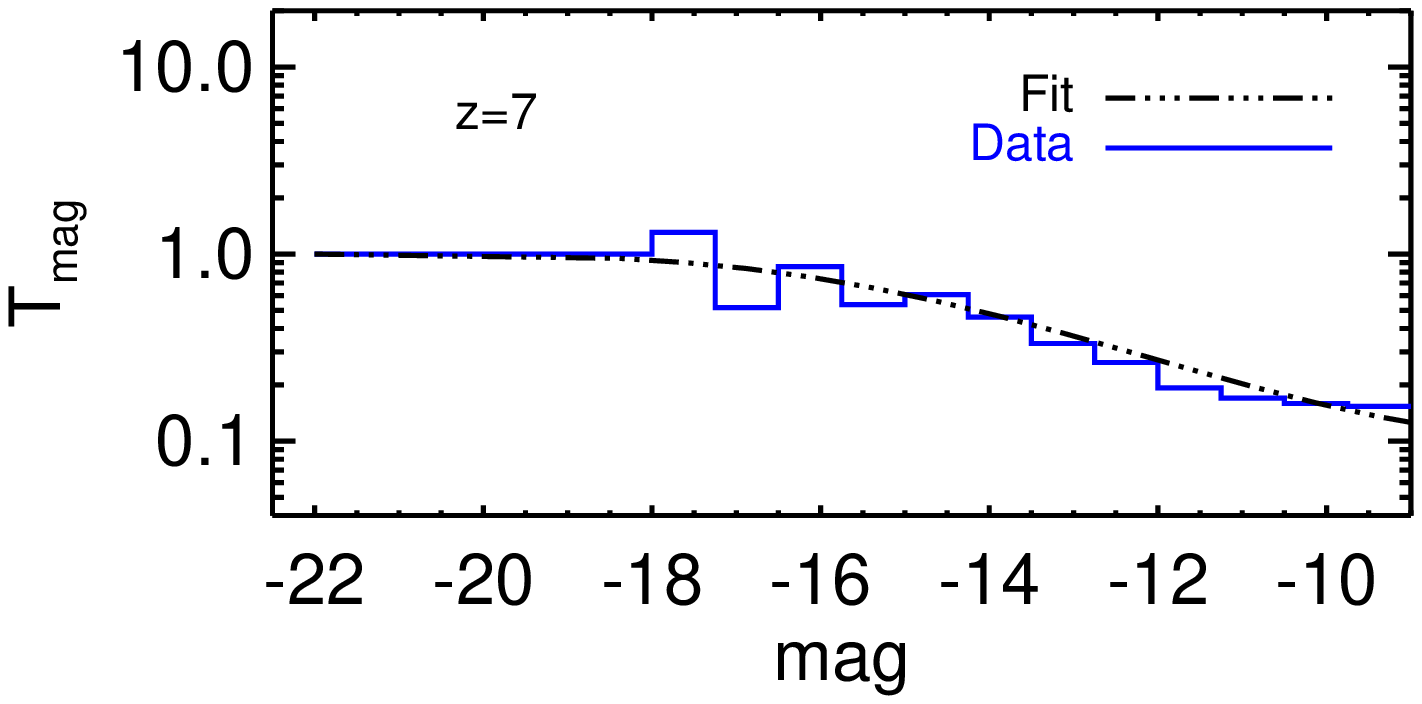}
\includegraphics[width=0.33\textwidth]{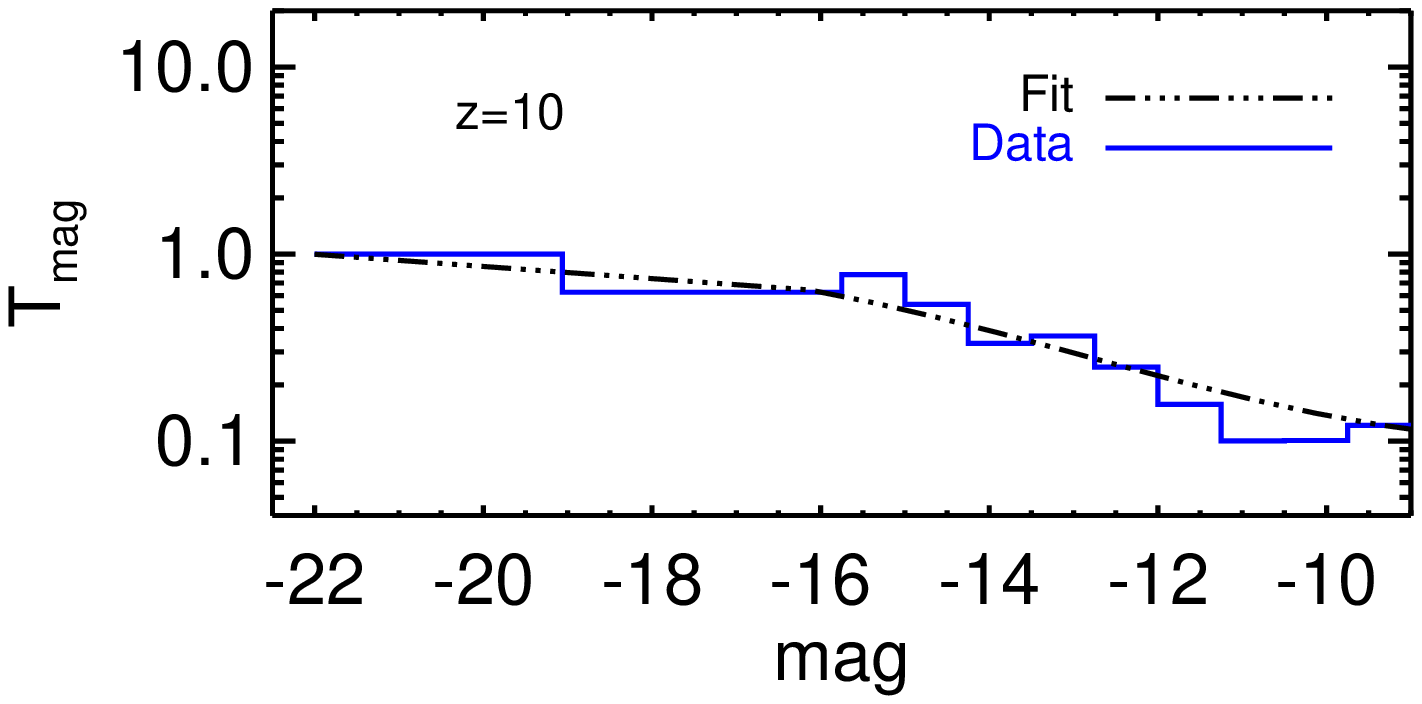}
\includegraphics[width=0.33\textwidth]{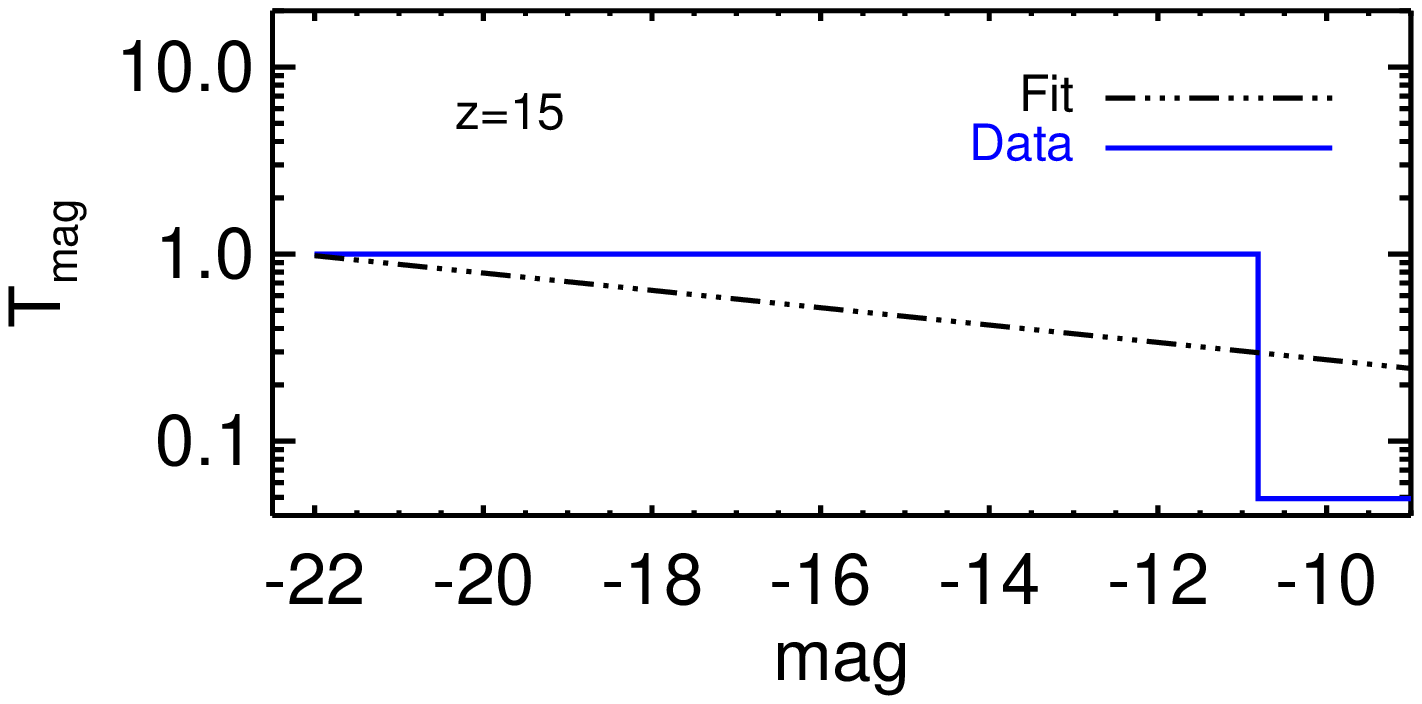}
\vspace{-0.3cm}
\caption[]{
  Upper panels show predicted UV luminosity functions at $z=7$ (left), $z=10$ (center) and $z=15$ (right) for CDM (solid lines) and WDM (dotted lines). The collection of observations (magenta squares) at $z=7$ includes available data at redshift $z\sim 6-8 $ according to \cite{Bouwens2007, Bouwens2011, Bouwens2014arXiv, McLure2010, Oesch2012}. Data and upper limits (magenta triangles) at $z=10$ are from \cite{Bouwens2014arXiv}.
Lower panels report the ratio between CDM and WDM luminosity functions (solid histograms) and the corresponding fit (dot-dashed lines) at different redshifts, as explained in the text (in case of missing data in the bright end $T_{\rm mag}=1$ has been assumed).
}
\label{fig:LF7}
\end{figure*}

\subsection{Observable signatures}\label{subsect:LF}
We conclude our study by discussing possible observational quantities that could be affected by the nature of dark matter during the cosmic dawn.
Besides metal distributions (Figure~\ref{fig:Zdistributions}), additional observables can be used to test CDM and WDM.

Interesting observational constraints might come from luminosity functions (LF) at high redshift.
UV magnitudes for galaxies at $z>6$ are available from a number of authors \cite[e.g.][]{McLure2010, Oesch2012, Bouwens2014arXiv} and can be tested against the expectations of numerical simulations.
In order to retrieve the simulated LF, the composite spectral energy distribution (SED) is assigned to each cosmic structure at different redshifts, according to its formation time, stellar mass, SFR, $Z$ and IMF by means of precomputed tables \cite[][]{BC2003}.
When dealing with multiple stellar populations (i.e. PopIII and  popII-I), as in the present case, it is necessary to distinguish the corresponding IMFs and yields for different generations, as well.
By following such {\it caveats} the simulation outputs can be converted into a luminosity distribution, displayed in Figure~\ref{fig:LF7}, as a function of UV magnitudes \cite[here the AB systems is adopted;][]{Oke1983}.
In the left panel, data points around redshift $z\sim 7$ are collected from \cite{Bouwens2007, Bouwens2011, Bouwens2014arXiv, McLure2010, Oesch2012} and compared to our theoretical expectations for both CDM and WDM.
The trend of the observed LF roughly agrees with a rather steep slope of $\beta \sim -2$ \cite[as also inferred by observations, see e.g.][]{Dunlop2013, Bouwens2014arXiv, Duncan2014arXiv} down to magnitudes of $\sim -15$ in both CDM and WDM scenarios.
At very bright luminosities a clear cut-off is well visible, as a consequence of the exponential decay of the dark-matter distribution at large masses.
These latter usually have larger $M_\star$ and higher SFR, hence, {\it cet. par.}, they can produce more luminous emissions.
Numerical results are in line with such behaviour, although rare big structures are missing in the simulated volumes, due to the finite size of our cosmological boxes.
\\
It is clear that at $z=7$ WDM and CDM are observationally indistinguishable, because data points lie in the very bright end of the distribution.
Therefore, only additional data samples at magnitudes fainter than $\sim -18$ might be able to break the degeneracy in the near future.
In particular, predicted spreads between CDM and WDM LFs suggest a spread in the abundance of luminous objects of 1 dex at magnitudes fainter than $\sim -12$.
We stress the differences in the two models by plotting the residuals in different magnitude bins. This allows us to better see the discrepancy at the faint end and to derive a useful fitting formula that can be employed to get estimates of the WDM effects at different $z$.
\\
We define the ratio between the WDM ($ \phi_{\rm WDM} $) and CDM ($ \phi_{\rm CDM} $) luminosity functions as
\begin{equation}
\label{eq:Tmag}
T_{\rm mag} = \frac{\phi_{\rm WDM}}{ \phi_{\rm CDM}}.
\end{equation}
This can be understood in view of its similarity to the transfer function of the WDM matter power spectrum.
We fit the simulated data (histograms) in the lower panels of Figure~\ref{fig:LF7}, with the following functional form:
%
%
\begin{equation}
\label{eq:fit}
T_{\rm mag}^{\rm Fit}(z) =1 - \beta\exp{ \left\{ - \left[\frac{mag}{mag_{\star}(z)} \right]^\gamma \right\} }
\end{equation}
with $\beta = 0.91$, $\gamma = 6$ and $mag_\star$ a reference ({\it break}) magnitude depending on redshift according to the relation
\begin{equation}
\label{eq:magstar}
mag_{\star}(z) = -16 \left( \frac{1+z}{10} \right)^{0.2}.
\end{equation}
The values of $mag_\star$ for the interested $z$ result $mag_\star(7) \simeq -15.3$, $mag_\star(10)\simeq -16.3$ and $mag_\star(15)\simeq -17.6$.
At magnitudes below $ mag_{\star} $ WDM and CDM tend to coincide, while at magnitudes larger than $ mag_{\star} $ there is a departure of WDM luminosity distributions from the CDM ones.
In the figure, the fitting formula is shown by black dot-dashed lines. The fit is quite satisfying although very high-$z$ it cannot be easily verified, because of poor statistics.
\\
At redshifts as high as $z=10$, the trends are still similar and highlight the need to go to magnitudes larger than $\sim -14$ to have departures of a factor of $\sim 2$ and to asses dark-matter nature via visible structures.
Around $z=15$ only CDM haloes can host gas molecular cooling, collapse, fragmentation and following star formation processes that could determine luminous emissions.
Instead, consistently with our previous discussions, emission from first WDM haloes results still cosmologically negligible, due to the retarded build-up of primordial stellar mass (Figure~\ref{fig:halomasses}) and paucity of early star forming episodes (Figure~\ref{fig:sSFR15}).
This means that a scenario dominated by a thermal relic with masses around $\rm 3~keV$ could be discriminated by a standard CDM cosmology by (future) high-redshift observations.

\begin{figure}
\vspace{-0.3cm}
\includegraphics[width=0.48\textwidth]{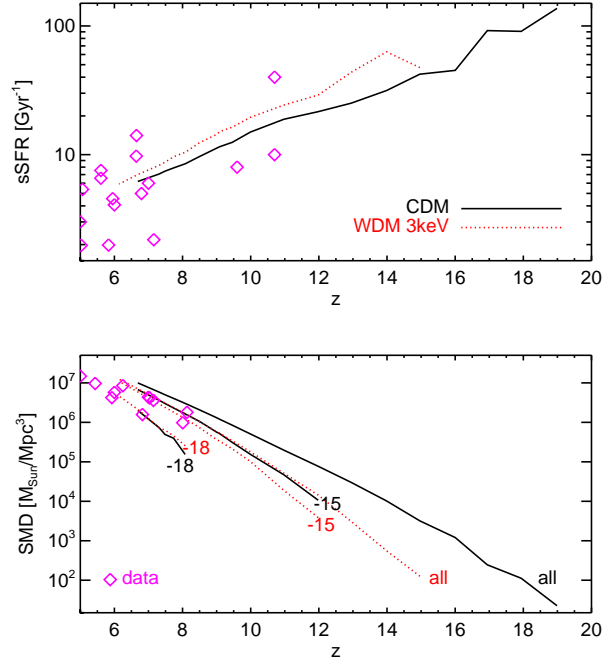}
\vspace{-0.5cm}
\caption[]{
  Redshift ($z$) evolution of the mean cosmic sSFR in $\rm Gyr^{-1}$ (upper panel) and of the stellar mass density (SMD) in  $\rm M_\odot/Mpc^3$ (lower panel) for both CDM (solid lines) and WDM (dotted lines).
As indicated by the labels, SMDs are computed:
(i) by considering {\it all} the structures in the simulated samples;
(ii) by considering a UV magnitude cut-off of $-18$;
(iii) by considering a UV magnitude cut-off of $-15$.
The collection of data points (magenta diamonds) for the sSFR is based on
\cite{Duncan2014arXiv, Stark2013, Bouwens2012, Gonzalez2012, Reddy2012, Zheng2012, Coe2013}. SMD data are from \cite{Labbe2010, Gonzalez2011, Stark2013,Duncan2014arXiv}.
}
\label{fig:SMD}
\end{figure}

In this respect, precious tools are offered by the analysis of the evolution of the stellar mass density (SMD) and the mean cosmic sSFR as functions of redshift.
The top panel of Figure~\ref{fig:SMD} shows the values attained by the mean sSFR between $z\sim 5$ and $z\sim 20$ for both CDM and WDM.
As a reference, recent observational constraints \cite[see][]{Duncan2014arXiv, Stark2013, Bouwens2012, Gonzalez2012, Reddy2012, Zheng2012, Coe2013} are also plotted.
Both trends are consistent with high-$z$ data and with previous CDM simulations at these epochs \cite[see discussion in e.g.][]{BiffiMaio2013}, however, they also highlight the more bursty nature of WDM structures with respect to the CDM counterparts (consistently with e.g. Sect.~\ref{subsect:SFactivity}).
\\
A more evident discrepancy is visible in the SMD evolution in the lower panel of Figure~\ref{fig:SMD}.
To fairly compare the predictions of our numerical simulations with observations, the SMD is computed both by considering all the formed objects and by considering observationally relevant magnitude cut-offs.
In particular, we show the SMDs corresponding to bright objects with UV magnitudes $\le -18$ and to fainter objects with UV magnitudes $\le -15$.
Moving the cut-offs towards higher thresholds leads to resulting behaviours that quickly reproduce the global one.
Data points are available only for $z<8$, a regime in which the CDM and WDM models roughly agree with observations, almost irrespectively of the considered magnitude thresholds.
At earlier epochs, there are more substantial differences that reach almost a factor of $\sim 10$ at $z\sim 12$ and a factor of $\sim 100$ at $z\sim 15$.
Therefore, future high-$z$ observations might be very useful to give more quantitative constraints on the scenarios analized here.

%


\section{Discussions and Conclusions}\label{Sect:conclusions}


We have investigated the effects of WDM on structure formation and compared them to predictions from the standard CDM scenario.
We have performed N-body hydro simulations including a detailed chemical treatment following $e^-$, H, H$^+$, H$^-$, He, He$^{+}$, He$^{++}$, H$_2$, H$_2^+$, D, D$^+$, HD, HeH$^+$ \cite[e.g.][]{Yoshida_et_al_2007, Maio2007, PM2012}, metal fine-structure transitions \cite[][]{Maio2007, Maio2009} and stellar evolution from popIII and popII-I stars \cite[][]{Tornatore2007, Maio2010, Maio2011b} according to individual metal yields (for He, C, N, O, Si, S, Fe, Mg, Ca, Ne, etc.) and lifetimes.
\\
In general, primordial structure formation in WDM haloes results delayed, because of the smaller amount of power at large $k$ modes with respect to the CDM scenario.
This determines a minor number of collapsing objects and substructures by 1 dex or more, a consequent suppression of molecular cooling and a global delay of (popIII and popII-I) star formation and metal pollution at early epochs.
These WDM induced effects are relevant for the birth of the first stars and galaxies, since the lack of star formation activity in smaller objects at $z\sim 10-20$ makes more difficult to produce ionizing photons capable of reionizing the Universe around $z\sim 6-10$.\\
Our conclusions were obtained by relying on some assumptions that could possibly affect the final results.
In particular, an important parameter is the assumed critical metallicity for stellar population transition that is subject to uncertainties of a few orders of magnitude, that constrain it in the range $\sim 10^{-6}-10^{-3}\,Z_\odot$.
Recent studies \cite[e.g.][]{Maio2010} have shown that, given the efficient metal spreading events at early times, such uncertainties for $Z_{\rm crit}$ would provoke only minor modifications to the resulting star formation history.\\
Similarly, our ignorance on the details of the pristine popIII IMF does not result to be crucial, because even in the most extreme cases popIII-dominated epoch remains confined in the first $\sim 10^7-10^8\,\rm yr$ from the onset of star formation and its contributions to the cosmic SFR rapidly drops of a few orders of magnitude by $z\sim 10$ \cite[e.g.][]{MaioIannuzzi2011}.
\\
Metal yields for stars of different masses and metallicities are periodically revisited and improved, but, despite the progresses in estimating individual abundances of different heavy elements more and more precisely, the overall amounts of the main metals ejected seem to roughly agree.
\\
Beyond dark-matter nature, further competitive effects at early times can additionally play a role.
For example, careful analysis of small-scale perturbations in primeval epochs \cite[][]{SZ1970, PressVishniac1980, HuSugiyama1996, TseliakhovichHirata2010, Fialkov2014} seem to suggest that primordial gas streaming motions could slightly lower the abundance of primordial gas clumps and delay star formation of some $\sim 10^7\,\rm yr$ on $\sim \rm Mpc$ scales \cite[e.g.][]{Maio2011a}.
However, the implications of WDM are found to be much more prominent.
\\
Alternative models suggest effects that are much smaller than the ones caused by WDM.
This is the case for non-Gaussianities that are constrained quite tightly at $f_{\rm NL}$ around unity and for which the implications on the star formation history would be negligible \cite[][]{MaioIannuzzi2011, Maio2011cqg, MaioKhochfar2012, Maio2012}.
\\
Dark-energy or quintessence scenarios also predict different trends for gas clouds and molecular evolution at early epochs \cite[e.g.][]{Maio2006} but, although strongly model-dependent, they rarely reach the consequences expected from WDM.
\\
Currently, available high-$z$ data are not sufficient to rule out WDM with thermal relic masses around 3 keV and statistical analysis of early faint objects would be needed.
Hopefully, in the next decades advanced instrumental facilities, such as SKA, MWA, LOFAR, PAPER, should be able to place concrete constraints on the cosmic dawn from radio observations of HI 21-cm emission and this could help unveil the nature of dark matter, as well.
\\
In short, our results can be summarized as follows.
\begin{itemize}
\item
  CDM and WDM power spectra differ by up to 2 dex at early times and show converging trends at $z<10$, although spreads of factor of a few persist in the whole first Gyr.
\item
  WDM structures below $\lesssim 10^{8}\,\rm M_\odot/yr$ are less abundant by at least 1 dex in the whole first Gyr, both in terms of dark content and in terms of baryonic content (gas and stars).
\item
  Simulated dark-matter mass functions in the first billion years reproduce fairly well theoretical expectations both for a CDM and for a WDM input power spectrum, without evidences of numerical fragmentation in WDM objects in the range of masses and redshifts of interest for the present work and independently from resolution.
\item
  Star formation in early WDM mini-haloes is suppressed and runaway molecular cooling becomes efficient only in larger objects at later times; as a result, the onset of star formation in the WDM scenario takes place with a delay of $\Delta z \sim 6$ compared to CDM.
\item
  SFR and sSFR distributions cover similar ranges in both dark-matter models, but the number of star forming galaxies in WDM haloes is about 10 times lower at $z=7$ and 100 times lower at $z=15$.
\item
  Typical gas fractions at early epochs are affected only up to a factor $\sim 2$ and their behaviour is dictated mainly by local thermodynamical processes.
\item
  Stellar-fraction distributions reflect the underlying dark-matter scenario by showing a {more remarkable} suppression of the number counts of more than 1 dex for $f_{\star}\sim 10^{-4}-10^{-2}$.
\item
  Physical correlations among visible quantities, such as SFR, sSFR, mass, $f_{\rm gas}$, $f_\star$ are led by local mechanisms and poorly influenced by the nature of the background dark-matter fluid.
\item
  Metallicity distributions are little affected by the background model and mainly led by stellar evolution.
\item
  Departures of the theoretical WDM luminosity functions from the CDM trends can be fitted by a simple formula depending only on magnitude and redshift.
\item
  WDM implications on structure growth are more dramatic than the ones from alternative models (e.g. non-Gaussian or dark-energy models) or from primeval high-order corrections (e.g. primordial streaming motions) and changes in the pristine IMF.
\item
  The theoretical lack of WDM objects during the epoch of reionization could help, in the near future, to pose serious constraints on the nature of dark matter by employing extensive searches of luminous objects at high $z$.
\item
  Detections of faint primordial (proto-)galaxies at $z\gtrsim 10$ could constrain a number of observables (LF, sSFR, SMD) in order to disentangle CDM from WDM.
\end{itemize}

In this paper, we have demonstrated that the early structure formation regime is a potentially promising environment to shed light on the very small-scale properties of dark matter via its physical implications on cosmological structures and the epoch of the cosmic dawn.
Thereby, the first billion years are an optimal tool to discriminate between CDM and WDM scenarios by means of current and next-generation instrumentation, such as 
SKA\footnote{http://www.skatelescope.org},
MWA\footnote{http://mwatelescope.org/index.php/science}, 
LOFAR\footnote{http://www.lofar.org}, 
HERA\footnote{http://reionization.org}
and possibly JWST\footnote{http://www.jwst.nasa.gov/}.


\section*{acknowledgments}
We acknowledge the anonymous referee for his/her detailed comments which significantly improved the presentation of this work.
UM acknowledges support from a Marie Curie fellowship by the European Union Seventh Framework Programme (FP7/2007-2013), grant agreement n. 267251.
MV is supported by the ERC-StG ``cosmoIGM'' and by INFN IS PD51 ``indark''.
We acknowledge the NASA Astrophysics Data System for the bibliographic~tools.


\bibliographystyle{mn2e}
\bibliography{bibl.bib}

\label{lastpage}
\end{document}